\def\feii{Fe~{\sc II}}
\def\fei{Fe~{\sc I}}

\documentclass[structabstract]{aa}  

\usepackage{graphicx}
\usepackage{graphics}
\usepackage{rotating}
\usepackage{psfig}
\usepackage{txfonts}
\usepackage{longtable,lscape}
\begin{document}

\title{Open clusters towards the Galactic center:\\ chemistry and dynamics}
\subtitle{ A VLT spectroscopic study of  NGC6192, NGC6404, NGC6583\thanks{Based on observations obtained at the ESO
Based on spectroscopic optical observations made with ESO telescope VLT, program: ID: 083.D-0682, Title: The Galactic radial metallicity gradient in the inner disk} }
\author{Laura Magrini\inst{1}, 
Sofia Randich \inst{1},   
Manuela Zoccali\inst{2},
Lucie Jilkova\inst{3,4}
Giovanni Carraro\inst{3,5},
Daniele Galli\inst{1}, 
Enrico Maiorca\inst{6,7},
Maurizio Busso\inst{6,7}}
%\offprints{L. Magrini}

\institute{
INAF--Osservatorio Astrofisico di Arcetri, Largo E. Fermi, 5, I-50125 Firenze, Italy\\
\email{laura--randich--galli@arcetri.astro.it}
\and
Departamento de Astronom\'ia y Astrof\'isica, Pontificia Universidad Cat\'olica de Chile, Av. Vicuna Mackenna 4860, Casilla 306, Santiago 22, Chile\\
\email{mzoccali@astro.puc.cl}
\and
ESO, Alonso de Cordova 3107, Vitacura, Santiago de Chile, Chile\\
\email{gcarraro-- ljilkova@eso.org}
\and
Department of Theoretical Physics and Astrophysics, Faculty of Science,
Masaryk University, (Kotl\'{a}\v{r}sk\'{a} 2, CZ-611 37) Brno, Czech
Republic
\and
Dipartimento di Astronomia, Universit\'a di Padova, Vicolo Osservatorio 3, I-35122, Padova, Italy
\and
Dipartimento di Fisica, Universit\'a di Perugia, via Pascoli, 06123 Perugia, Italy\\
\email{busso--maiorca@fisica.unipg.it} 
\and
Istituto Nazionale di Fisica Nucleare, sezione di Perugia, via Pascoli, 06123 Perugia, Italy \\ 
}

\date{}
\abstract
{In the framework of the study of the Galactic metallicity gradient and its time evolution, we present new high-resolution spectroscopic observations obtained with FLAMES and the fiber link to UVES at VLT 
of three open clusters (OCs) located within $\sim$7~kpc from the Galactic Center (GC): NGC~6192, NGC~6404, NGC~6583. We also present  new orbit determination for all OCs with Galactocentric distances (R$_{\rm{GC}}) \leq$8~kpc and metallicity from high-resolution spectroscopy.  }
{We aim to investigate the slope of the inner disk metallicity gradient as traced by OCs and at discussing its implication on  the chemical
evolution of our Galaxy. }
{We have derived memberships of  a group of evolved stars for each clusters, obtaining a sample of 4, 4, and 2 member stars
in NGC~6192, NGC~6404, and NGC~6583, respectively. 
Using standard LTE analysis we derived stellar parameters and abundance ratios for the iron-peak elements Fe, Ni, Cr, and for the $\alpha$-elements Al, Mg, Si, Ti, Ca. 
We calculated the orbits of the OCs currently located within 8~kpc from the GC, and discuss their implication 
on the present-time radial location. }
{The average metallicities of the three clusters are all oversolar: 
[Fe/H]= $+0.12\pm0.04$ (NGC~6192), $+0.11\pm0.04$ (NGC 6404), $+0.37\pm0.03$ (NGC 6583).
They are in qualitative agreement with their Galactocentric distances, being all internal OCs, and thus  
expected to be metal richer than the solar neighborhood. The abundance ratios of the other elements over iron [X/Fe] 
are consistent with solar values. }
{The clusters we have analysed, together with other OC and Cepheid data, confirm a steep gradient in the inner disk, a signature of an evolutionary rate different than
in the outer disk.   
%Young clusters (age$<$1~Gyr) in the inner disk (R$_{GC}$$<$8~kpc) demonstrate  that  metallicity does not increase 
%up to extremely oversolar values, but reaches a saturation due to the complete consumption of the 
%gas and to the maximum iron yield.  
} 

\authorrunning{Magrini, L. et al.}
\titlerunning{Chemistry and dynamics of OCs in the inner disk}
\maketitle
\section{Introduction}

Theoretical models of chemical
evolution allow us to understand  the processes involved in the formation and evolution of galaxies, 
and in particular of the Milky Way (MW). 
However they depend on many variables, which include: 
the star formation rate (SFR), the initial mass function (IMF), radial 
inflows and outflows of primordial and/or pre-enriched gas, the star formation law 
and possible thresholds in the gas density for the formation of stars. 
To decrease the number of {\em degrees of freedom} they need many strong observational constraints, as 
those provided by the radial metallicity gradient and its evolution with time. 

In spite of many observational and theoretical efforts, various questions remain open 
even in our Galaxy, where  plenty of observations based on 8m-class telescopes of stellar populations 
of different ages are available.  
We are far from obtaining a firm result about the shape of metallicity gradient in particular in the inner disk and 
at large Galactocentric distances, and about its time evolution.

Contrasting results about the time evolution appear from the comparison of abundances of  planetary nebulae of different 
ages (cf. Maciel et al.~\cite{maciel07}; Stanghellini \& Haywood \cite{stanghellini10}), Cepheids 
(e.g., Pedicelli et al.~\cite{pedicelli09}; Andriewsky et al.~\cite{an02}), HII regions (e.g., Rood et al.~\cite{rood07}) 
and blue supergiant stars (e.g., Luck et al.~\cite{luck89}; Daflon \& Cunha \cite{daflon04}).  

In this framework, the determination of the chemical composition of stars belonging to 
open clusters (OCs) represents probably the best tool to study the behaviour of iron and 
other elements across the Galactic disk. 
OCs are indeed populous, coeval groups of stars with similar chemical
composition; they are located all over the Galactic disk
and span large ranges of ages and metallicities;
furthermore, the estimate of their age and distance
is affected by much smaller uncertainties than those of field stars.

The first determination of the Galactic metallicity gradient with the use of OC metallicities 
was performed by Janes (\cite{janes79}). 
He based his study on a sample of 
OCs covering the Galactocentric distance (R$_{\rm GC}$)
range 8-14~kpc and on metallicities 
([Fe/H]) estimated from photometry,  deriving a gradient 
of $-0.075\pm$ 0.034~dex~kpc$^{-1}$. 
Several subsequent studies were performed using a variety
of techniques and larger samples of clusters
(Friel \& Janes \cite{friel93}; Friel \cite{friel95}; Twarog et al. \cite{twarog97}; Carraro et al. \cite{carraro98};
Friel et al. \cite{friel02}; Chen et al. \cite{chen03}), all agreeing on a negative
slope between $-0.07$ and $-0.1$~dex~kpc$^{-1}$. 

More recently, with the advent of high-resolution spectrographs on 8m-class telescopes, it has become
possible to derive more secure [Fe/H] estimates, and also to extend metallicity determinations
to OCs in the outer disk, beyond 15~kpc from the Galactic center 
(Carraro et al.~\cite{carraro04}; Carretta et al.~\cite{carretta04}; 
Yong et al.~\cite{yong05}; Sestito et al.~\cite{sestito06}, \cite{sestito08}). 
These studies have confirmed the steep slope in the gradient for  7~kpc$<R_{\rm GC}<$11~kpc
and  have shown that the distribution becomes flatter for Galactocentric distances ($R_{\rm GC}$) above 11--12~kpc
(cf. Magrini et al. \cite{m09}, hereafter M09, but see  Pancino et al.~\cite{pancino10} for a different view). 
 
However, in spite of the significant progresses so far achieved, a very interesting radial region, 
R$_{\rm GC} < 7$~kpc, has been neglected during the past spectroscopic  studies of OCs. 
This radial region  is particularly interesting  because from Cepheid metallicities, several authors 
(e.g.,  Andrievsky et al.~\cite{an02}; Luck et al.~\cite{luck06}; Pedicelli et al.~\cite{pedicelli09}),  have shown that there is a further change of slope of the gradient. 
From $\sim$4 to $\sim$7-8 kpc (values vary slightly in different papers) the gradient is much steeper than in 
the outer regions. The value of the slope of the very inner ($R_{\rm GC}<$7~kpc) gradient for iron  varies from $-$0.13 to $-$0.15~dex kpc$^{-1}$. 
The questions are if this holds true also for OCs,  if this change of slope is time-dependent (and OCs can tell us this because 
they span a large age range), and 
where the metallicity stops increasing due to the presence of the Galactic bulge. 
So-far only one cluster  at a distance from the GC smaller than 7~kpc has an available metallicity
determination (M11, R$_{\rm GC}$=6.86~kpc, age=200~Myr, [Fe/H]$=+$0.17 --
Gonzales \& Wallerstein \cite{gonzales00}). 

Several studies and chemical evolution models have tried to explain the changes of slope in the abundance gradient, among them, e.g.
{\em i)} the study of dynamical effect of corotation resonance (located close to the solar Galactocentric distance) by L{\'e}pine et al.~(\cite{lepine01})
which is assumed  to be the 
main cause of the formation of the bimodal radial distribution of metallicity ; 
 {\em ii)} the model by M09 
in which the inner disk is formed inside-out by the rapid dynamical collapse of the halo and  the outer disk
is formed at a lower rate, from a 'reservoir' of gas;
{\em iii)} the model by Colavitti et al. (\cite{colavitti09}) where 
the inside-out disk formation and the density threshold for  star formation 
are both necessary ingredients to reproduce the change of slope;
{\em iv)} the model by Fu et al. (\cite{fu09}) where the  star formation efficiency is 
inversely proportional to the distance from the galaxy center. 

The prediction of these models are also different for what concerns the inner gradient.  
They all reproduce one to a few changes of slope in radial gradient but at different $R_{\rm GC}$:
for example, the model  adopted in the study by L{\'e}pine et al.~(\cite{lepine01}) has a flattening of the gradient for  $R_{\rm GC}>R_{\odot}$, while the model by  
M09 produces a gradient with an outer plateau ($R_{\rm GC}>11-12$kpc), and a very steep gradient for $R_{\rm GC}<7$kpc. 
Finally, the best model by Colavitti et al. (\cite{colavitti09}) has a more pronounced gradient in the inner disk for 
$R_{\rm GC}<7-8$kpc, while the models by Fu et al. (\cite{fu09}) show a change of slope in the inner Galaxy for $R_{\rm GC}<2-3$kpc. 

The aim of this paper is to extend the sample of OCs at $R_{\rm GC}<$7~kpc, 
thus providing constraints suitable to disentangle the roles of different model parameters
in inducing changes into the chemical gradient and its slope
as, e.g.,  the importance the dynamical 
collapse of the halo in the inner regions with respect to the accretion due to infall from the intergalactic medium
and from mergers. 

The present paper is structured as follows: in Sect.~\ref{sec_tar} the three OCs under analysis are presented. 
In Sect.~\ref{sec_obs} we describe the observations and data reduction. 
In Sect.~\ref{sec_abu} we show the abundance analysis, while in Sect.~5 we present our results and compare them  
 with the inner disk giant stars. whereas in in Sect.~6   
we discuss their implication in the shape of the metallicity gradient and in the chemical evolution of the Galaxy. 
In Sect.~\ref{sec_orbit} we present orbit calculations and their consequences for the metallicity gradient. 
Finally, in Sect.~\ref{sec_sum} we give our summary and conclusions.

\section{Target clusters}
\label{sec_tar}

Our  sample  includes three of the most internal open 
clusters (age $\leq$1~Gyr) whose evolved stars can be studied with a ground-based 8m-class 
telescope. They are NGC~6192, NGC~6404, and NGC~6583. 

Several photometric studies of NGC 6192 are present in the literature. 
These studies derived,  however, different  reddening values 
and, consequently, discrepancies in ages and distances exist.
The reddening values in the literature range from E(B-V) = 0.26 (Kilambi \& Fitzgerald \cite{kilambi83}; King \cite{king87}) 
to 0.68 (Kjeldsen \& Frandsen \cite{kj91}), with the most recent estimate by  Paunzen et al. (\cite{paunzen03})
giving  E(B-V) = 0.54. The related estimates of age range from 
$\sim$1~Gyr to 0.09~Gyr, and those for the distance to the Sun from $\sim$1~kpc to 1.7~kpc. 
Paunzen et al. (\cite{paunzen03}) estimated also the metallicity of the cluster:   [Fe/H] = $-$0.10 $\pm$ 0.09. 
Loktin et al. (\cite{loktin01}) revised the original data of Kjeldsen \& Frandsen (\cite{kj91}), and 
determined  a reddening of E(B-V) = 0.64, a distance of 1.5~kpc and an age of 0.13~Gyr.  
Five stars of NGC 6192 (Nos. 9, 45, 91, 96 and 137) have CORAVEL radial velocities in the narrow 
range -8.8 km s$^{-1}$ $<$ Vr $<$ -6.4 km s$^{-1}$, appearing thus 
to be red giant members of NGC 6192 (Clar{\`i}a et al. \cite{claria06}).
Clar{\`i}a et al.~(\cite{claria06}) also estimated the age (0.18~Gyr), the distance (1.5~kpc), and the $\it DDO$ abundance index
finding  a cluster metallicity [Fe/H] = +0.29$\pm$0.06.
\begin{figure*}
  \centering
\includegraphics[angle=-90,width=16cm]{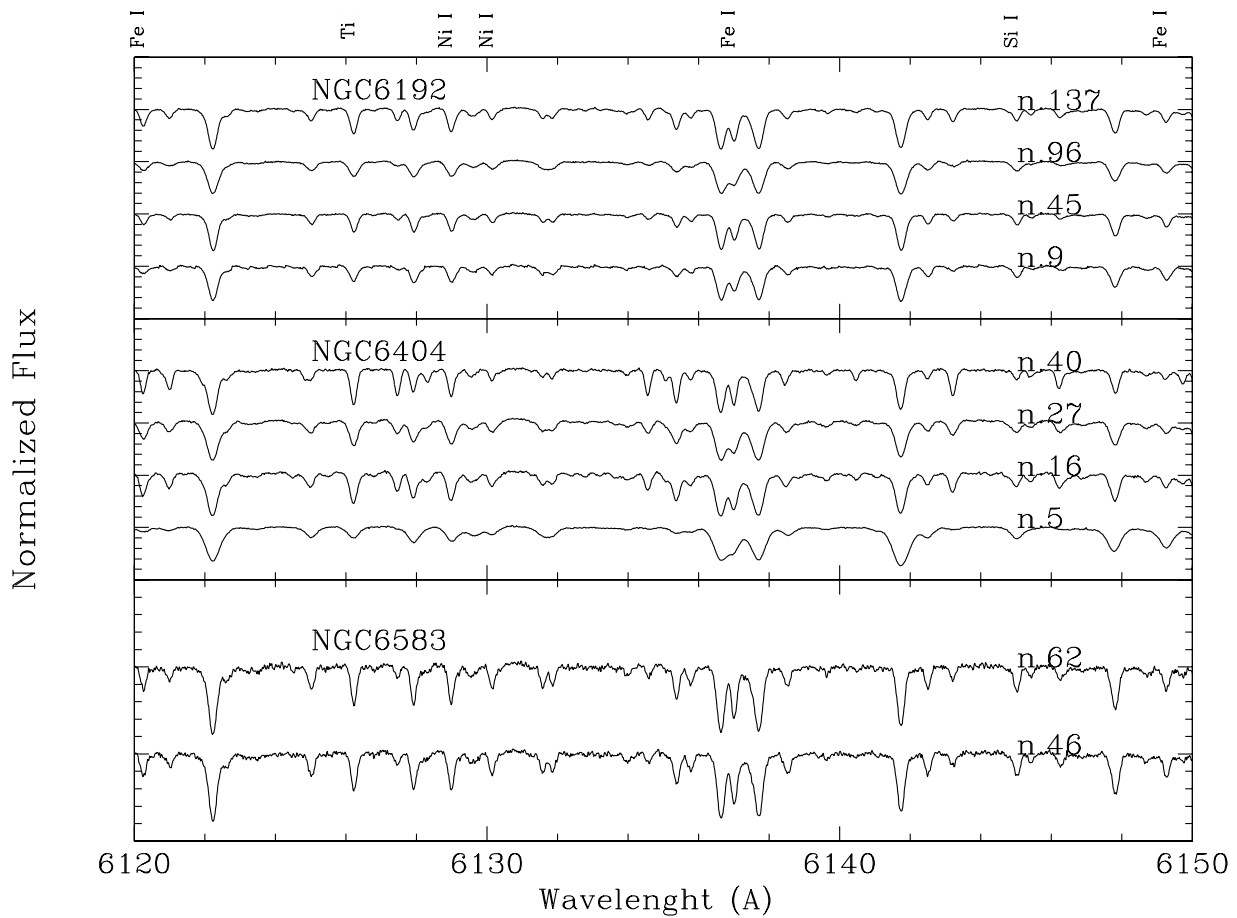}
    \caption{A region of the normalized spectra of member stars in NGC~6192, NGC~6404, NGC~6583.}
\label{Fig_spectra}%
\end{figure*}

NGC~6404 and NGC~6583 have been studied by Carraro et al. (\cite{carraro05}) with CCD photometry. 
From  isochrone fitting they derived the main parameters of the two clusters. 
They found that NGC~6404 is 0.5 Gyr old and located at a distance of 1.7~kpc from the Sun towards the 
Galactic Centre direction, while NGC 6583 is 1.0 Gyr old and at 2.1 kpc.
E(B-V) values are 0.92$\pm$0.05 and  0.51$\pm$0.05  for NGC~6404 and NGC~6583, respectively. 
For both clusters, solar metallicity isochrones provide a reasonable fit across the whole CMDs. 
No radial velocities are available for the stars of these clusters. 

In Table~1 we report the  cluster parameters and references. We also calculated the 
Galactocentric distance,  R$_{\rm GC}$, adopting a distance of the Sun from the Galactic centre of 
8.5~kpc.

\begin{table}
\begin{center}
\label{tab_par}
\caption{Cluster parameters from the literature. }
\begin{tabular}{ccccccc}
\hline
\hline
Cluster & Age 		& [Fe/H] 		& R$_{\rm GC}$ 	& $d$ 	& $E(B-V)$ & Ref. \\
        	 & (Gyr)        	&                 	&  (kpc)      		& (kpc) &       		&        \\
 \hline
% NGC~6005 &~1.2 &~----             &~6.46 &~2.7 &~0.45 \\
 NGC~6192 &~0.18 	&$+0.29$ 		&~7.1 			&~1.5 	&~0.64 & a\\
 NGC~6404 &~0.5 	&~solar          &~6.8 			&~1.7 &~0.92  & b\\
 NGC~6583 &~1.0 	&~solar          &~6.4			&~2.1 &~0.51  & b\\
 \hline\hline
\end{tabular}\\
\end{center}
a) Clar{\`i}a et al.~(\cite{claria06})\\
b) Carraro et al.~(\cite{carraro05})\\
\end{table}

\section{Observations and data reduction}
\label{sec_obs}

The three OCs  were all observed with the multi-object instrument 
FLAMES on VLT/UT2 (ESO, Chile; Pasquini et al. 2000). The   fibers with UVES 
allowed us to obtain high-resolution spectra (R=47~000), for Red Giant Branch (RGB) stars, for red supergiant 
stars and for stars in the clump.
The clusters were observed in Service mode in April, July, and August 2009, using a single 
configuration for each cluster. The cross disperser CD3 was used   resulting in a wavelength range  $\sim$4750-6800 \AA. 
A log of observations (date, centre of field, date, observing time of each exposure, grating, number of exposures, number of stars) is given in Table 
\ref{obslog}. The spectra were reduced by ESO personnel using the dedicated pipeline. 
We analyzed the 1-d, wavelength-calibrated spectra using standard IRAF\footnote{IRAF is distributed by the National Optical Astronomy Observatory, which is operated by the Association of Universities for Research in Astronomy (AURA) under cooperative agreement with the National Science Foundation.} packages and the 
{\sc SPECTRE} package by Sneden et al. (Fitzpatrick \& Sneden \cite{fitz87}).
One fiber for configuration was used to register the sky value, but the correction was negligible. 
In Figure \ref{Fig_spectra} we show a region of the spectra of stars in NGC~6192, NGC~6404, NGC~6583 (only member stars, see below). 
In Figures Fig.~\ref{Fig_cdm_6192}, Fig.~\ref{Fig_cdm_6404}, and  Fig.~\ref{Fig_cdm_6583} we present the color-magnitude diagrams of the three clusters. 
The stars observed spectroscopically are marked with filled (red) circles. Stars which turn out to be non-members are marked also with a cross.

\begin{figure}
  \centering
   \includegraphics[angle=270,width=8cm]{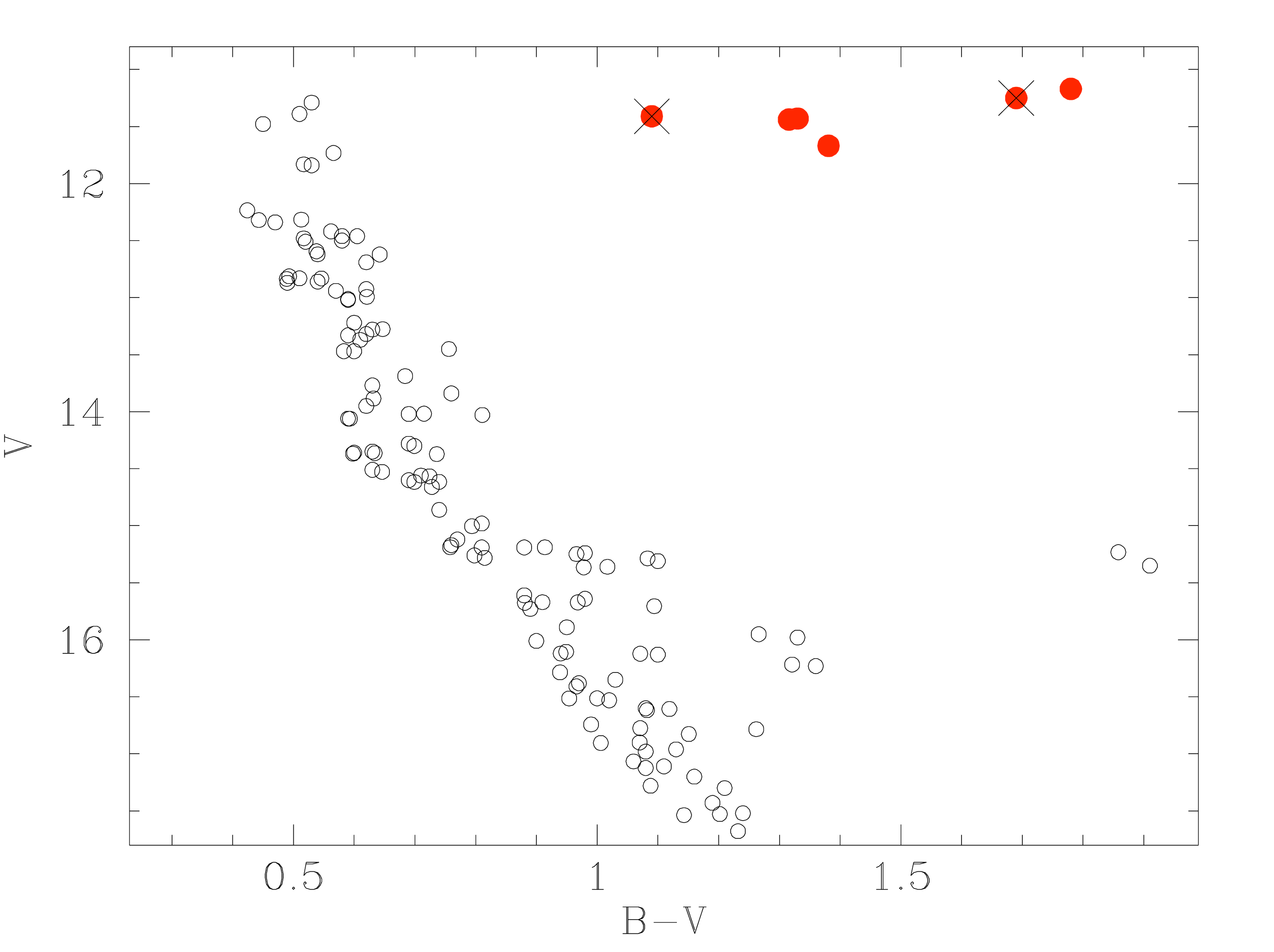}
    \caption{Colour-magnitude diagram  of NGC6192. The observed stars are marked with (red) filled circles. Stars which turn out to be non-members are marked also with a cross.  
    The photometry is from Clar{\`i}a et al.~(\cite{claria06}) }
\label{Fig_cdm_6192}%
\end{figure}
\begin{figure}
  \centering
   \includegraphics[angle=270,width=8cm]{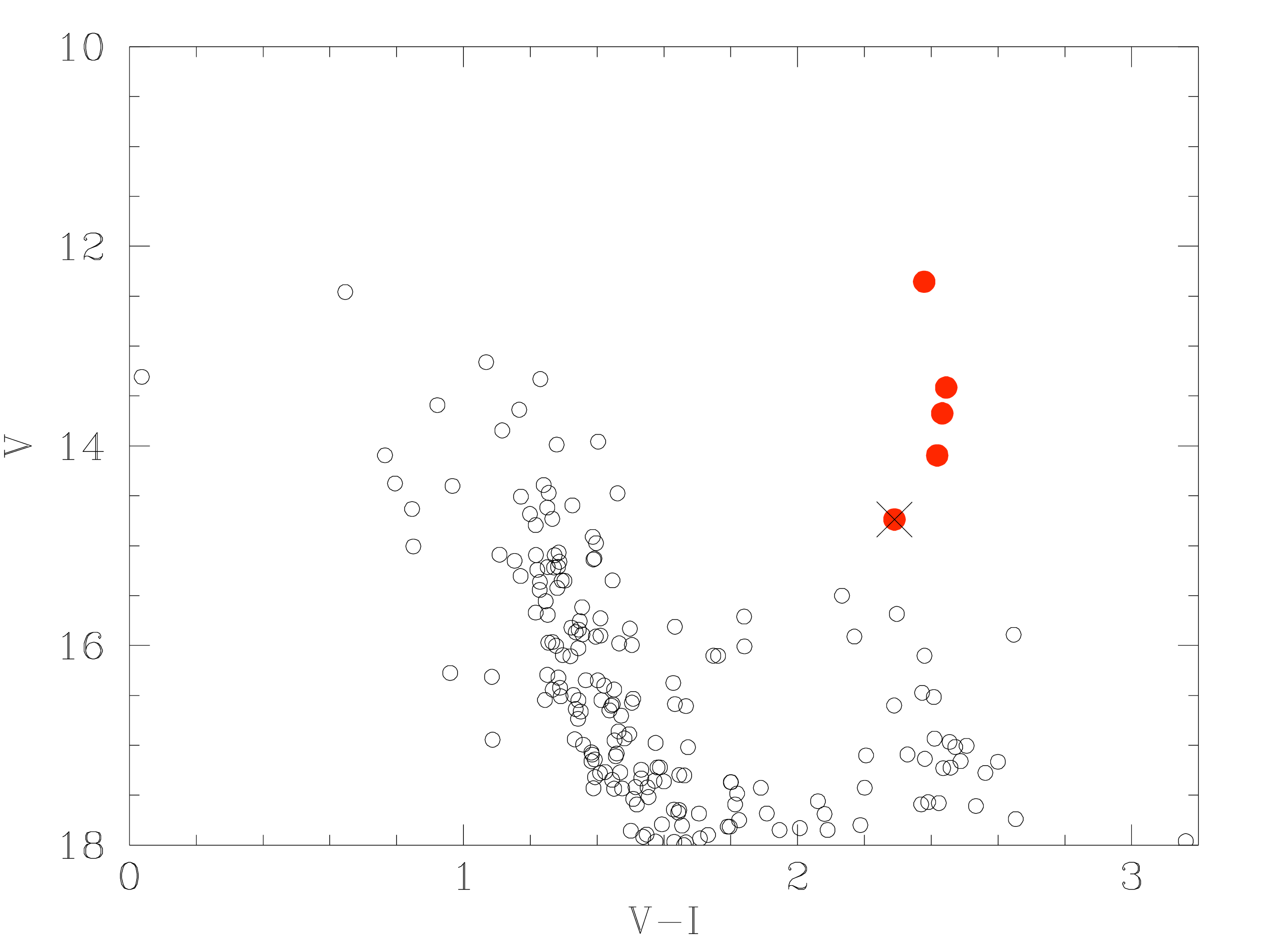}
    \caption{Colour-magnitude diagram of NGC6404. Symbols as in Fig.~\ref{Fig_cdm_6192}. The photometry is from Carraro et al.~(\cite{carraro05}).  }
\label{Fig_cdm_6404}%
\end{figure}
\begin{figure}
  \centering
   \includegraphics[angle=270,width=8cm]{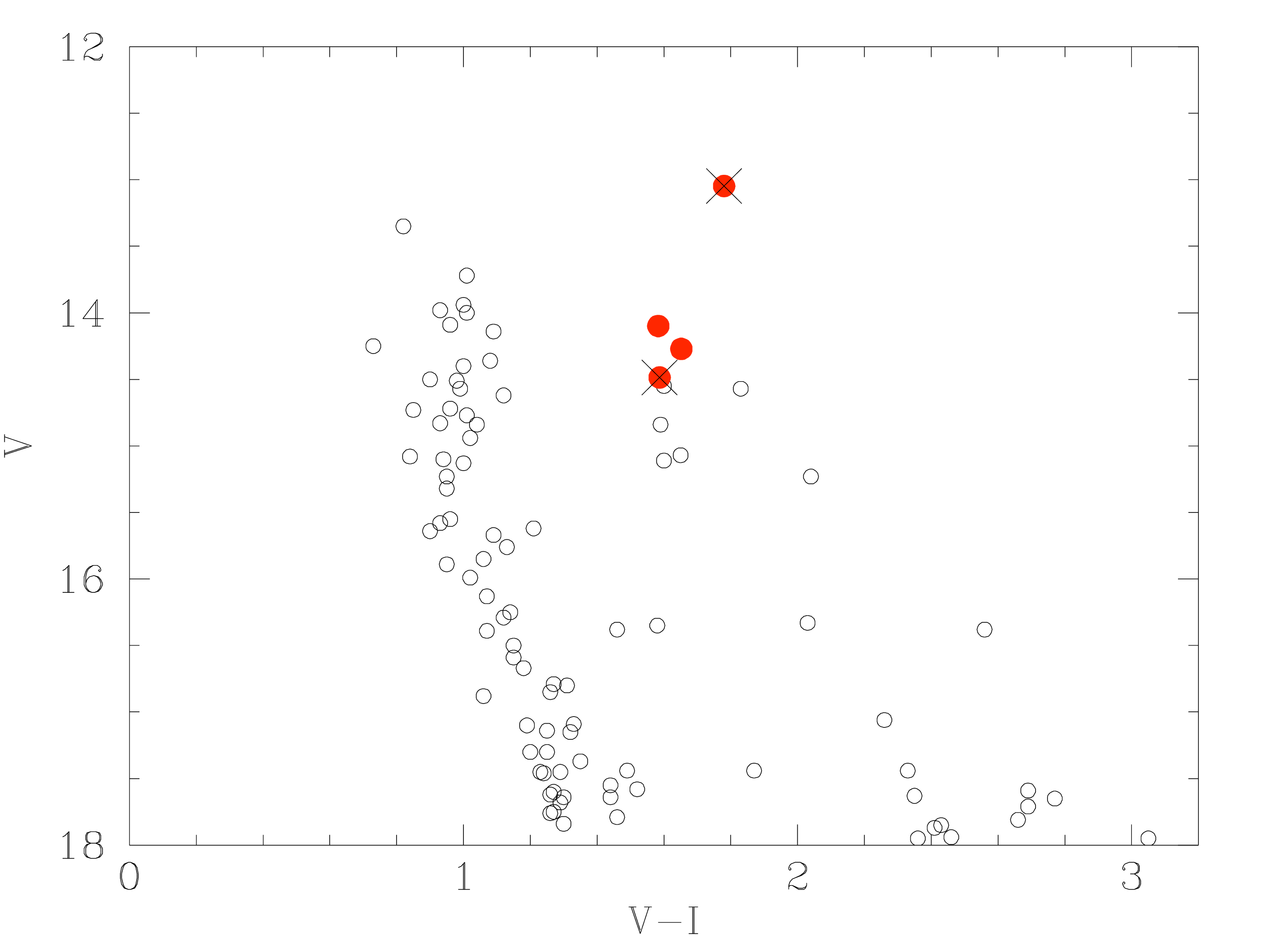}
    \caption{Colour-magnitude diagram of NGC6583. Symbols as in Fig.~\ref{Fig_cdm_6192}. The photometry is from Carraro et al.~(\cite{carraro05}).  }
\label{Fig_cdm_6583}%
\end{figure}
\begin{table*}
\centering \par \caption{Log of the observations.\label{obslog}} 
\begin{tabular}{lccccccc} 
\hline\hline 
Cluster   & \multicolumn{2}{c}{Centre of field}             				    &Date             &Exptime &CD & No. of exp. & No. of stars \\ 
               & RA & Dec    &yyyy-mm-dd &(s)          &        & tot. &\\ 
               & \multicolumn{2}{c}{J2000.0}   & &        &        & &\\ 
\hline 
NGC6192 &16 40 50& -43 22 00  &2009-04-09  &2775 & CD3 & 1 &6\\ 
NGC6404 &17 33 37& -33 14 48  &2009-04-09  &2775 & CD3 & 4 &5\\ 
			&			&		    &2009-04-11 &2775  & CD3 &  &\\ 
			&			&		    &2009-08-06 &2775  & CD3 &  &\\
			&			&		    &2009-08-06 &2775  & CD3 &  &\\  
NGC6583 	&18 15 48& -22 08 00&2009-04-09  &2775 & CD3 & 5 &4\\ 
			&			&		    &2009-07-28 &2775  & CD3 &  &\\
			&			&		    &2009-08-20 &2775  & CD3 &  &\\ 
			&			&		    &2009-08-20 &2775  & CD3 &  &\\ 
			&			&		    &2009-08-20 &2775  & CD3 &  &\\
 \hline\hline
\end{tabular} 
\end{table*} 

First of all, we measured the Radial Velocities (RVs) using {\sc RVIDLINES} on several tens of metallic 
lines on the individual spectra. Using the RVs measured on the single lines, the  task  {\sc RVIDLINES} 
produces an average radial velocity with its {\it rms} dispersion. We measured the radial velocity 
of each star on each individual spectrum, correcting them for the contribution of  the different 
heliocentric velocities during each night of observations.
The uncertainty associated to the RVs of  each star, corrected for the heliocentric component,  
is obtained from the rms value, after averaging estimates from different 
exposures; it turns out to be   0.5 km s$^{-1}$.  

Finally we combined the spectra obtained during the different exposures with the task {\sc SCOMBINE}. 

In Tables  \ref{tab_n6192_s}, \ref{tab_n6404_s}, \ref{tab_n6583_s}, we show  the available photometry 
of the observed stars,  the signal to noise ratio (S/N), the RVs, corrected for the heliocentric velocity, and the membership 
derived  on the basis of the RVs (M stands for ``member'', NM for ``non member''). 
For NGC6192 the identification numbers and UBV magnitudes  were taken from Clar{\'i}a et 
al.~(\cite{claria06}), for NGC6404 and NGC6583 the ID numbers and VI photometry are
from Carraro et al. (\cite{carraro05}).
For all clusters the JHK photometry is from the 2MASS catalogue (Skrutskie et al.~\cite{skrutskie06}).
There are previous radial velocity measurements for the evolved stars of NGC6192 (Clar{\`i}a et al. \cite{claria06}). 
We found good agreement 
with their values.  Their average velocity for  five member stars, including two 
binary stars (No. 91 and 96) is $-7.7\pm0.38$~km~s$^{-1}$. The average velocity of our member sample 
is   $-$8.9$\pm$0.7~km~s$^{-1}$ in good agreement with the average velocity computed for the same 
stars using the Clar{\`i}a et al.'s measurements,  $-$8.4$\pm$0.5~km~s$^{-1}$. 
 
Previous radial velocity measurements are not available for NGC6404 and NGC6583. 
From our spectroscopy, the average velocities of member stars are  10.6$\pm$1.1~km~s$^{-1}$ (4 stars)
and -3.0$\pm$0.4~km~s$^{-1}$ (2 stars) form NGC6404 and NGC6583, respectively.

For our further analysis we thus exclude stars Nos. 4 and 274 of NGC6192, No. 72 of NGC6404, and 
Nos. 12 and 82 of NGC6583 on the basis of their non-membership.

\begin{table*}
\begin{center}
\caption{Data of observed stars in NGC6192.  }
\begin{tabular}{cccccccccccll}
\hline\hline
NGC 6192&&&&&&&&&&&&	\\ 
\hline
Star & RA$_{2MASS}$  & Dec$_{2MASS}$  & V & B-V & J$_{2MASS}$ & H$_{2MASS}$ & K$_{2MASS}$ & RV & S/N & Notes \\
 		&                            &                              &   &                 &                                                 &                         & &@6500\AA                & (km s$^{-1}$)   & \\
\hline
 0004   & 16 40 45.63&   -43 24 07.8  & 	11.588  & 1.334  &9.094 & 8.345 &  8.187   &170& +46.3 & NM \\
 0009   & 16 40 40.81&   -43 22 59.4  & 	11.430  & 1.510  &8.478  &7.883  & 7.645   &170 & -9.4  & M\\
 0045   & 16 40 24.21&   -43 20 08.0  & 	11.743  & 1.546  &8.674  &7.986  &7.797    &160& -9.1  & M\\
 0096   & 16 40 16.46&   -43 22 37.6  & 	11.306  & 1.502  &8.277  &7.619  &7.371    &140 & -9.7  & M\\
 0137   & 16 40 11.42&   -43 24 00.8  & 	11.223  & 1.758  &7.776  &7.002  &6.757    &140& -7.6 & M\\
 0274   & 16 40 35.69&   -43 27 22.2  & 	11.300  & 1.650   &7.948  &7.142  &6.888    &180 & +2.1 & NM\\
\hline
\hline
\end{tabular}
\label{tab_n6192_s}
\end{center}
\end{table*}

\begin{table*}
\begin{center}
\caption{Data of observed stars in NGC6404. }
\begin{tabular}{ccccccccccll}
\hline\hline
NGC 6404&&&&&&&&&&&	\\ 
\hline
Star & RA$_{2MASS}$  & Dec$_{2MASS}$  & V & V-I & J$_{2MASS}$ & H$_{2MASS}$ & K$_{2MASS}$ & RV & S/N & Notes \\
 		&                            &                              &   &         &        &                                                &         & @6500\AA                                  & (km s$^{-1}$) & \\
\hline
   5  &17 39 30.86   &  -33 14 52.3  & 12.35 	& 2.379 &  7.892  & 7.139  &6.803  &120 &+9.7&M \\ 
  16  &17 39 43.55  &   -33 14 06.6 &  13.42 &  2.445 &  8.716  &7.748  &7.427  &110 & +9.8&M \\
  27  &17 39 43.46  &   -33 15 14.0  & 13.67 	& 2.433 &  8.989  & 8.061  &7.685  &150 &+10.3&M \\ 
  40  &17 39 37.24  &   -33 14 10.1  & 14.09 	& 2.418 &  9.471  & 8.390  &8.098  &120 &+12.8&M \\  
  72  &17 39 34.92  &   -33 14 39.3  & 14.74 	& 2.293 &  10.279& 9.385  &9.056  &100 &-42.9&NM \\  
\hline
\hline
\end{tabular}
\label{tab_n6404_s}
\end{center}
\end{table*}

\begin{table*}
\begin{center}
\caption{Data of observed stars in NGC6583.  }
\begin{tabular}{ccccccccccll}
\hline\hline
NGC 6583&&&&&&&&&&&	\\ 
\hline
Star & RA$_{2MASS}$  & Dec$_{2MASS}$  & V & V-I & J$_{2MASS}$ & H$_{2MASS}$ & K$_{2MASS}$  & RV & S/N & Notes \\
 		&                            &                              &   &         &        &                         &                                 & @6500\AA                                & (km s$^{-1}$)   & \\
\hline
  12  &   18 15 50.04  &   -22 08 21.4   &    13.05	&1.788  &       9.663   &    8.885    &  8.663  &  90  &+44.5&NM \\ 
  46  &   18 15 51.13  &   -22 07 26.4   &    14.10 	&1.583   &     10.986  &   10.331   & 10.170 &  100 & -3.4&M \\   
  62  &   18 15 51.23  &   -22 08 28.0   &    14.27 	&1.652   &     11.056   &  10.379   &10.186  &  80& -2.6 &M \\  
  82  &   18 15 47.54  &   -22 08 31.5   &    14.49 	&1.587   &     11.105   &  10.381   & 10.198 &  100  &+8.7&NM \\        
\hline
\hline
\end{tabular}
\label{tab_n6583_s}
\end{center}
\end{table*}

\section{Abundance analysis}
\label{sec_abu}

The method of analysis is very similar to that adopted in the papers by Randich et al.~(\cite{randich06}) and 
Sestito et al. (\cite{sestito06}, \cite{sestito08}), 
based on measured equivalent width (EWs) and on the use of the code MOOG.  In this paper 
the analysis of chemical abundances was carried out with the version 2002 of the spectral program MOOG (Sneden 1973) and using model atmospheres by Kurucz (\cite{kurucz93}). Like all the commonly used spectral analysis codes, MOOG performs a local thermodynamic equilibrium (LTE) analysis.  

\subsection{Line list and equivalent widths}
Spectral lines to be used for the analysis of Fe~I, Fe~II and other elements (Mg, Al, Si, Ca, Ti, Cr, Ni) 
both in the Sun and in our giant stars were selected from the line list of Gratton et al. (\cite{gratton03}) and 
Randich et al.~(\cite{randich06}). The total  list includes 161 and 14 features for Fe~I and Fe~II, respectively, and several 
features of Al I, Mg I, Ca I, Si I, Ti I, Cr I, Ni I. 
In order to have a strictly differential analysis, we selected the same spectral range for the Sun and for our 
giant stars. 
%For this reason,  iron lines in the bluest part of the spectra, i.e. at wavelengths 
%bluer than $\sim$5500 \AA, are not included in our list to avoid complications 
%due to crowding and continuum tracing in the red giant star spectra.

We used the task {\sc eq} in SPECTRE to normalize small portions of the spectrum ($\sim$40 \AA) and to 
interactively measure the EWs of the spectral lines by Gaussian fitting. 
Strong lines ($EW > 150$ m\AA) have  been discarded, since they could be saturated and they 
are critically sensitive to the microturbulence value, thus  
a more detailed treatment of damping would be necessary to fit their line wings. We also discard faint lines with 
$EW < 20$ m\AA\, because of the large uncertainties in their measurements.    
The values of EWs are available in electronic Tables A.1-3, where the first two columns list the wavelengths and element, and the others show the corresponding EW for each star.

\subsection{Atomic parameters and solar analysis}
We adopted the  oscillator strengths ($\log gf$) of Sestito et al. (\cite{sestito06}), 
with exception of  those of Al obtained by Randich et al. (\cite{randich06}) from inverse solar analysis. 
Radiative and Stark broadening are treated in a standard way in MOOG. 
We used the  Uns\"{o}ld (\cite{unsold55})  approximation  for collisional broadening since this choice
does not affect the differential analysis with respect to the Sun as discussed by Paulson et al. (\cite{paulson03}).
In addition,  very strong lines that are most affected by the treatment of damping have been excluded from our analysis.
The major difference with the previous studies of Sestito and collaborators is indeed in the different treatment of 
damping. Here we have chosen a more homogeneous treatment for all lines   	 
using  the Uns\"{o}ld approximation with respect to the procedure by  Sestito et al., who adopted two different 
types of damping coefficients, those of Barklem et al. (\cite{barklem00}), when available, 
otherwise the  coefficients of Uns\"{o}ld, multiplied by an enhancement factor. 
In order to check the compatibility of the two approaches,  we considered the EWs of two stars 
in NGC~6253 (No. 023498 and 
024707 in  Table~4 of Sestito et al.~\cite{sestito07}) with stellar parameters similar to those of our stars. 
We recalculated [Fe/H] using our procedure, finding for No.~023498 and 
No.~024707 the [Fe/H] values $+$0.30 and $+$0.40, respectively.  
They compare well with the values quoted in the original paper, i.e., $+$0.32 and $+$0.39, respectively.

The first step is to derive the solar abundances of Fe and other elements, in order to fix the zero 
points of the abundance scale  to minimize errors in the results. 
The solar spectrum was obtained with the same UVES setup of our observations. 
The line list for the Sun is available in electronic form (Table A.4). 
The table includes  wavelengths, name of the element, the excitation potential EP, $\log gf$, and measured EW in the solar spectrum. 
We computed the elemental abundances for the Sun by adopting the following effective temperature, surface gravity, and microturbulence velocity: $T_{\rm eff}$ = 5770 K, $\log g=4.44$, and $\xi=1.1$ km~s$^{-1}$. 
In our analysis neither $\log$ n(Fe  I) vs. EW nor $\log$ n(Fe  I) vs. EP  show any trend, implying 
that both the assumed microturbulence and effective temperature are correct. 
$\log$ n(Fe~I) and $\log$ n(Fe~II) are consistent within the errors indicating also a correct 
choice of the gravity value. 
Output solar abundances are listed in Table \ref{tab_sun_abu} together with those from Anders \& Grevesse (\cite{anders89}) 
that are used as input in MOOG.
Our \fei\ abundance is slightly lower than that  measured by Anders \& Grevesse (\cite{anders89}). The origin of this discrepancy 
is probably due to the adopted $\log gf$, but it would not affect the further analysis since it is strictly differential.

\begin{table}%t3 
\caption{Element abundances for the Sun in our analysis (T$_{\rm{eff}}$=5770, $\log g=4.44$, $\xi$=1.1), and in Anders \& Grevesse (\cite{anders89}, AG89).}
\label{tab_sun_abu} %
\centering \small \begin{tabular}{lrcc}
 \hline \hline 
El. & No. & $\log n$(X)$_{\odot}$ & $\log n$(X)$_{\odot}$ \\ 
&lin. & Our & AG89 \\ \hline 
Mg~{\sc i} & 2 &7.56 $\pm$ 0.03 &7.58  \\ 
Al~{\sc i} & 2 &6.49 $\pm$ 0.03 &6.47  \\ 
Si~{\sc i} & 10 &7.58 $\pm$ 0.02 &7.55  \\ 
Ca~{\sc i} & 9 &6.29 $\pm$ 0.03 &6.36  \\ 
Ti~{\sc i} & 14 &4.93 $\pm$ 0.03 &4.99  \\ 
Cr~{\sc i} & 7 &5.67 $\pm$ 0.03 &5.67  \\ 
Fe~{\sc i} &59 &7.47 $\pm$ 0.03 &7.52 \\ 
Fe~{\sc ii} & 9 &7.49 $\pm$ 0.02 &--  \\ 
Ni~{\sc i} & 22 &6.24 $\pm$ 0.03 &6.28  \\ 
\hline\hline
\end{tabular} \end{table}

\subsection{Stellar parameters}

\begin{table*}
\caption{Photometric and spectroscopic stellar parameters and [Fe/H] abundances. }
\begin{center}
\begin{tabular}{ccccccccccc}
\hline
\hline
Star & T$_{eff(B-V)}$ & T$_{eff(V-K)}$& log $g_{phot}$    &T$_{eff spec}$&  log $g_{spec}$ & $\xi$ &[Fe/H]&$\sigma$1 & No lines\\
        & (K)        		   & (K)                  &                  				&		(K)					&								&  km s$^{-1}$       &&&\\
\hline
NGC6192 &      &               & & &&&&\\
\hline
  9 &  5070 & 5190 & 2.34 & 5050& 2.30 & 1.75 & $+$0.19 & 0.07&59 (8)\\
 45 &  5000 & 5000 & 2.38 &5020 & 2.55 & 1.60&$+$0.08  &0.08&119 (9)\\
 96 &  5090 & 5010 & 2.23 &5050 &2.30 & 2.10 &$+$0.13  & 0.10& 77 (8)\\
137 &  4580 & 4500 & 1.90 & 4670 & 2.10 &1.80& $+$0.07& 0.08&91 (9)\\
\hline
Average [Fe/H]	&				&		&			&		  &	&	& $+0.12$ &0.04       &       \\
\hline\hline
NGC6404 &      &               &    & &&&&&\\
\hline
   5 &  -				& 4300& 1.96 &5000 &1.0 &2.60   &$+0.05$& 0.09 &63 (3)\\
 16 &  -				& 4060& 2.08 &4450 & 1.65 & 2.10 &$+0.07$&0.09&67 (8)\\
27 &  -				& 4060& 2.17 &4400 & 1.40 & 1.80 &$+0.20$ & 0.09& 47 (4)\\
 40 &  -			 	& 4060 & 2.29 &4250 & 2.30 & 1.40 &$+0.11$ & 0.10& 89 (5)\\
\hline
Average [Fe/H]	&				&		&			&		  &	&	& $+0.11$ &0.04       &       \\
\hline\hline
NGC6583 &      &               & & &&&&&   \\
\hline
 46 &  -				& 4625  & 2.90 & 5100 & 2.95 & 1.45 & $+$0.40 & 0.12 & 68 (8)\\
 62 &  -			 	& 4510  & 2.91 &5050   & 2.75 &1.45 & $+$0.34& 0.12& 71 (10)\\
\hline
Average [Fe/H]	&				&		&			&		  &	&	& $+0.37$ & 0.03      &       \\
\hline
\hline
\end{tabular}
\label{tab_stars}
\end{center}
\end{table*}

We initially estimate effective temperatures ($T_{\rm eff}$) and gravities 
from photometry. 
We used the B$-$V and V$-$K vs.  $T_{\rm eff}$ calibrations by Alonso et al.~(\cite{alonso99}) 
for the RGB stars of NGC6192, whereas we used V$-$K vs.  $T_{\rm eff}$ calibrations by Levesque et al.~(\cite{levesque06})
for the post-RGB stars of NGC6404 and NGC6583.  
Surface gravities were derived using the expression 
$$\log g=\log (M/M_{\odot})+0.4(M_{\rm bol}-M_{\rm bol{\odot}})+4 \log (T_{\rm {eff}}/T_{\rm {eff}\odot})+\log g_{\odot},$$ where $M$ is the mass and $M_{\rm bol}$ 
the bolometric magnitude (with $M_{\rm bol{\odot}}=4.72$), $T_{\rm eff}$ is derived as described above (with  
$T_{\rm {eff}\odot}$=5770 K), and $\log g_{\odot}$=4.44. 
The apparent magnitudes are transformed into absolute magnitudes using 
the  distances and reddening values in Table~1.
The bolometric corrections to be applied  to $M_{\rm V}$ in order to obtain   
$M_{\rm bol}$ are taken from Alonso et al.~(\cite{alonso99}) in the 
case of giant stars and from Levesque et al.~(\cite{levesque06})
for post-RGB stars.  
The masses were derived from the isochrones of Girardi et al. (\cite{girardi00})  
using  the estimated ages of each cluster (see Table~\ref{tab_par}). 
They are:
4.1 M$_{\odot}$ for NGC 6192 (age $\sim$ 0.18 Gyr), 
3.0 M$_{\odot}$ for NGC 6404 (age $\sim$ 0.5 Gyr), 
2.0 M$_{\odot}$ for NGC 6583 (age $\sim$ 1 Gyr).

The photometric $T_{\rm eff}$ and $\log g$ values are the starting values.  
For NGC6192, where two photometric estimates of $T_{\rm eff}$ are available, 
we used as initial value their average.   
We optimized  their  
values  during the spectral analysis, employing the driver 
{\sc ABFIND} in {\sc MOOG} to compute Fe abundances for the stars: the final $T_{\rm eff}$ estimate was chosen in order 
to eliminate possible trends in [Fe/H] vs. EP. 
The surface gravity was optimized by assuming the ionization equilibrium condition, i.e. $\log n(\rm {Fe}$  II)=$\log n(\rm {Fe}$  I). 
Then, if necessary $T_{\rm eff}$ was re-adjusted in order to satisfy both the ionization and excitation equilibria. 
The other stellar parameter is the microturbulence $\xi$ that was optimized by minimizing the slope of the 
relationship between \fei/H and the observed EWs. 

The abundance values  for each line were determined with the measured EWs and stellar parameters 
listed in Table~\ref{tab_stars}. 
Final abundances for each star and each element were determined as the mean abundance from the different lines. 
Due to the large number of lines available, a 2$\sigma$ clipping was performed for iron as the first step  before the 
optimization of the stellar parameters.
As already mentioned, [Fe/H] and [X/Fe] ratios for each star were determined differentially with respect to solar 
abundances listed in Table~\ref{tab_sun_abu}.

The derived photometric and spectroscopic stellar parameters are shown in Tab. \ref{tab_stars}: 
col. 1 presents the star ID, cols. 2-4 the photometric $T_{\rm eff}$ and $\log g$ values, and 
cols. 5-7 the spectroscopic $T_{\rm eff}$, $\log g$ and $\xi$ estimates. Finally, in cols. 8-9 the iron abundance 
and its dispersion are shown, whereas in col. 10 the number of \fei\ and \feii, in parenthesis, lines 
available for the spectroscopic analysis are presented.

\begin{figure}
  \centering
   \includegraphics[angle=0,width=8cm]{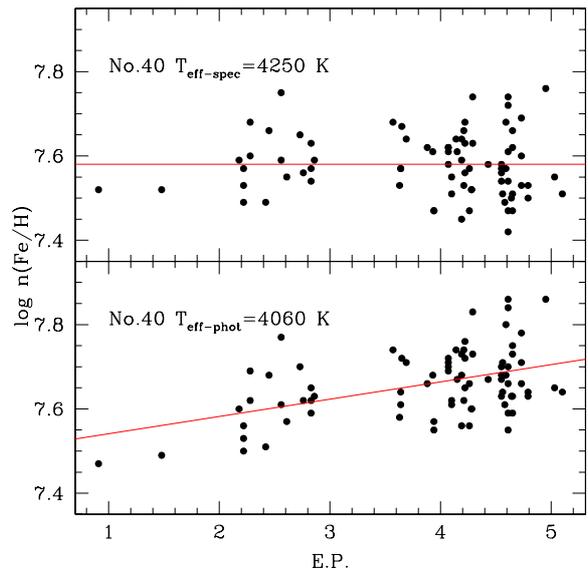}
    \caption{log {\em n}(Fe/H) vs. E.P. for star No.40 in NGC~6404 with two T$_{eff}$: 4060 K as derived from photometry 
    and 4250 K as derived from the fulfillment of excitation equilibrium with the spectroscopic analysis with MOOG.   
    The two continuous lines are the  mean least-square fits to the data performed by MOOG.   }
\label{Fig_temp_ep}%
\end{figure}

For NGC~6192 we found reasonable agreement between the photometric and spectroscopic parameters, 
while for NGC~6404 and NGC~6583 the differences are important. 
The origin of this discrepancy derives from the large uncertainty  on the reddening value and on the distance, 
due to the location of these clusters towards the GC. 
In fact, as far as $T_{\rm eff}$ estimates  of these two clusters are concerned, we note that in order to satisfy 
the excitation equilibrium  we had to consider $T_{\rm eff}$ values larger than those derived from the  photometric 
relations (see as an example Fig.~\ref{Fig_temp_ep}).
%, with average differences  
%of $\Delta$(${T_{\rm eff}}_{\rm {phot}}-{T_{\rm eff}}_{\rm {spec}}$)=-400$\pm$190 K 
%and 500$\pm$50 K for NGC~6404 and NGC~6583, respectively. 
These differences can be reduced by assuming that slightly higher values of the reddening 
with respect to those measured by Carraro et al. (\cite{carraro05}) through a photometric analysis. 
For both clusters, a value of $\Delta$A$_V$ 
larger than the one quoted in Table 1 by 0.60 would 
reduce the difference 
among photometric and spectroscopic T$_{\rm{eff}}$. 
%to 60$\pm$160 K and 20$\pm$70 K for NGC6404 and NGC6583, respectively.
The remaining differences in surface gravities can be attributed 
to several  factors, such as  errors in distance moduli.

\subsubsection{Evolved stars  in NGC~6404 and NGC~6583}
\begin{figure}
  \centering
   \includegraphics[angle=0,width=8cm]{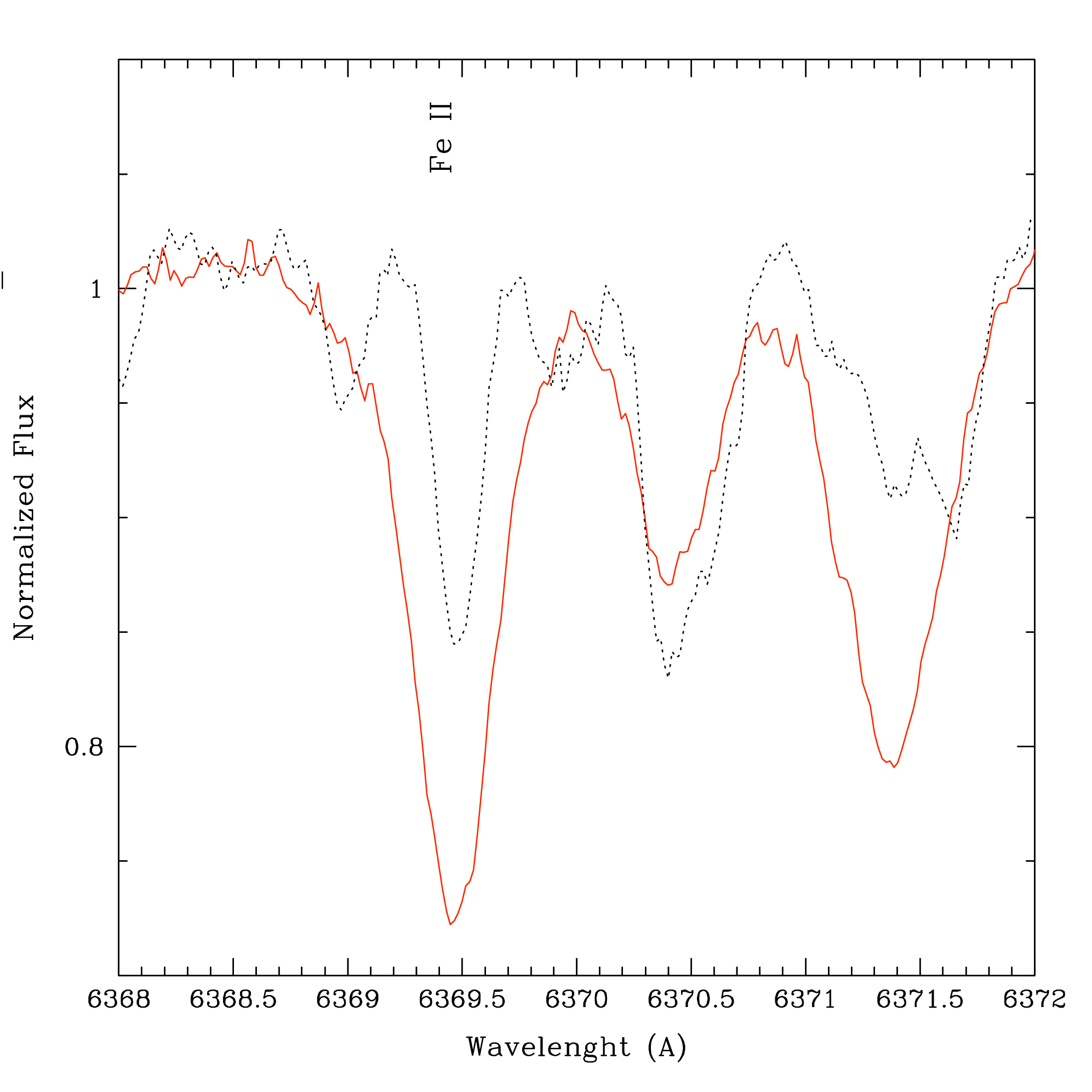}
    \caption{A region of the normalized spectra of member stars in NGC6404 centred in the [FeII] line at 6369.45 \AA.
    The continuous (red) spectrum is of star No.~5, which has a very low gravity, while the dotted line is the spectrum of star No.~45 in NGC~6192,
    having similar temperature and metallicity, but higher gravity.    }
\label{Fig_spectra_feii}%
\end{figure}

The criterion of selection of evolved stars to be spectroscopically investigated  is based on their location 
in the color-magnitude diagram.  Due to different distance and reddening of the three clusters 
we did not select stars in the same evolutionary phase in the three OCs. 
In particular, the high reddening towards NGC6404 and NGC6583 did not allow us to observe
red clump giant stars, which lie at V$>$16. 
The evolved stars we have selected in NGC6404 and NGC6583 for our spectroscopic analysis (see Figs.~\ref{Fig_cdm_6404} and 
\ref{Fig_cdm_6583}) 
are indeed in a latter evolutionary stage with respect to the RGB phase, moving towards the asymptotic giant branch phase. 
The most extreme example  is  star No.~5 in NGC6404 (in the color-magnitude diagram of Fig.~\ref{Fig_cdm_6404} this is the star with the lowest V magnitude). 
This star has an extremely low surface gravity combined with a high effective temperature. 
To show the reliability of our surface gravity measurement we plot in Fig.~\ref{Fig_spectra_feii}
a portion  of the normalized spectra located around a Fe~II line of star No.~5 of NGC~6404 and of a star with similar 
effective 
temperature but with higher gravity, No.~45 in NGC~6192. 
The EWs of Fe~II lines are inversely related to the surface gravity, thus lower gravity implies 
larger EWs of Fe~II. 
This is the case of star No.5 showing  very strong Fe~II lines, for which we measured $\log~g=1$ (see Table 7).  
Also the high microturbulence of star No.5 (2.6~km~s$^{-1}$) is in agreement with its advanced  evolutionary phase, 
being comparable with the microturbulence of  red supergiant stars ($\sim$3~km~s$^{-1}$, cf. Davies et al. \cite{davies09}). 

\subsection{Error evaluation}

As usually done, we considered three sources of errors in our analysis: 
{\em i)} errors in measurements of the EWs;
{\em ii)} errors in atomic parameters;  
{\em iii)} errors due to uncertainties in stellar parameters.

We estimated the errors in the following way. 
We consider the standard deviation ($\sigma$1) around the mean abundance derived by individual lines 
as representative of the errors in EWs. We show  the {\em rms} uncertainties in Table~\ref{tab_stars} for \fei\, and 
\feii\, and in Table~\ref{tab_abu} for the other elements.   
We consider as negligible the errors in atomic parameters because they are   minimized by the differential analysis  
performed with respect to the Sun. 
We estimated internal errors due to uncertainties in stellar parameters ($\sigma$2) by varying each parameter separately, 
while leaving the other two unchanged. Following the error analysis by Randich et al. (\cite{randich06}), we assumed random uncertainties 
of ${\pm} 70$ K, ${\pm} 0.15$ km s$^{-1}$, and ${\pm} 0.25$ dex in $T_{\rm eff}$, $\xi$ and $\log~g$, respectively.
We varied the stellar parameters of the warmest and coolest stars of our sample, No.~46 of NGC6583, and No.~40 of NGC6404, 
respectively.
The variations in the abundance ratios due to changes in the stellar parameters are shown in Table~\ref{errors}. 
Then we assumed, as an estimate of the error due  to the uncertainties on the stellar parameters, the maximum variation reported in Table~\ref{errors} for each element.

\begin{table}%t6 %
\centering \par \caption{Random errors due to uncertainties in stellar parameters.\label{errors}} 
\begin{tabular}{llll}\hline\hline 
\multicolumn{4}{c}{No.46~: $T_{\rm eff}$~=~5100~K, $\log~g$~=~2.95, $\xi=1.45~km~s^{-1}$}\\ \hline 
$\Delta$ & $\Delta T_{\rm eff} = \pm 70$ & $\Delta \log g=\pm 0.25$ & $\Delta \xi = \pm0.15$\\ 
			& (K) 						& dex 						& km~s$^{-1}$ \\ 
%5170 5030 3.20 2.70 1.30 1.60
\hline 
~[Fe~I/H] 	& 0.04/-0.05	 	& 0.02/-0.01	 	&  -0.07/0.08 \\ 	% 7.87 7.85/7.92 7.86/7.89  7.95/7.80 %nella table sono invertite, 
~[Mg/H] 	& 0.03/-0.03 		& 0.00/0.00 		&  -0.03/0.03 \\ 	% 7.90 7.88/7.93  7.90/7.90 7.93/7.87 sono +delta, -delta
~[Al/H] 	& 0.05/-0.04 		& 0.00/0.00 		&  -0.03/0.04 \\ 	% 7.00 6.97/7.04  7.01/7.00  7.04/6.97
~[Si/H] 	& 0.00/0.00 		& 0.05/-0.02 		&  -0.04/0.04 \\ 	% 7.80 7.83/7.79  7.78/7.85  7.84/7.77
~[Ca/H] 	& 0.07/-0.07	 	& 0.00/-0.05 		&  -0.08/0.07 \\ 	% 6.73 6.68/6.79  6.75/6.70  6.80/6.65
~[Ti/H] 		& 0.09/-0.09	 	& 0.00/0.00 		&  -0.07/0.08 \\ 	% 5.35 5.29/5.44  5.35/5.34  5.43/5.28
~[Cr/H] 	& 0.08/-0.07 		& 0.00/0.00 		&  -0.07/0.08 \\ 	% 5.88 5.83/5.95  5.88/5.87  5.95/5.81
~[Ni/H] 	& 0.03/-0.03 		& 0.04/-0.03 		&  -0.07/0.08 \\ 	% 6.68 6.68/6.74  6.65/6.72  6.76/6.61
\hline 
\multicolumn{4}{c}{No.~40: $T_{\rm eff}$~=~4300~K, $\log~g$~=~2.30, $\xi=1.4$~km~s$^{-1}$}\\ \hline 
$\Delta$ & $\Delta T_{\rm eff}~ =\pm 70$ & $\Delta \log g$~=~$\pm 0.25$ & $\Delta \xi=\pm~0.15$\\ 
& (K) & dex & km~s$^{-1}$ \\ \hline 
%4370 4230  2.60 2.1 1.63 1.33
~[Fe~I/H] 	& 0.02/-0.02 	&  0.01/-0.01 	& -0.05/0.07 \\ 
~[Mg/H] 	& 0.01/-0.01 	&  0.05/-0.05 	& -0.02/0.03 \\ 
~[Al/H] 	& 0.02/-0.02 	&  0.01/-0.01 	& -0.02/0.04 \\ 
~[Si/H] 	& 0.07/-0.06 	&  0.07/-0.07 	& -0.03/0.04 \\   %	7.74		7.68/7.81	7.81/7.66	7.71/7.78
~[Ca/H] 	& 0.07/-0.06 	&  0.00/-0.01 	& -0.08/0.08 \\ 	%	6.22		6.29/6.16	6.21/6.22	6.14/6.30
~[Ti/H] 		& 0.09/-0.08 	&  0.01/-0.02 	& -0.13/0.09 \\ %5.07		5.16/4.99	5.08/5.05	4.96/5.20
~[Cr/H] 	& 0.07/-0.05 	&  0.01/-0.01 	& -0.08/0.10 \\ 	%	5.65		5.72/5.60	5.66/5.64	5.57/5.75
~[Ni/H] 	& 0.03/-0.02 	&  0.07/-0.08 	& -0.07/0.08 \\ %6.43		6.41/6.46	6.50/6.35	6.36/6.51
\hline\hline 
\end{tabular} 
\end{table}

\section{Results}
\subsection{Iron}

\begin{figure}[!t]
\centering
\includegraphics[width=0.45\textwidth]{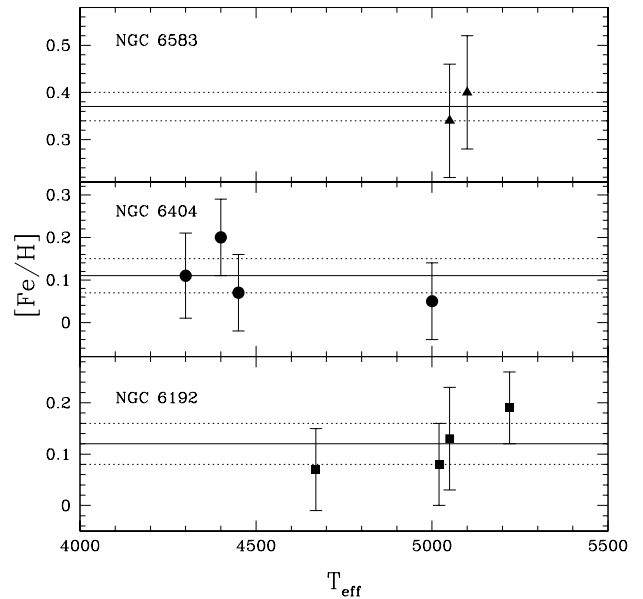} 
\caption{[Fe/H] versus the effective temperature in the studied member stars of the three clusters: NGC~6583 (filled triangles), 
NGC~6404 (filled circles), NGC~6192 (filled squares).  }
\label{Fig_fe_te}
\end{figure}

[Fe/H]  abundances and {\em rms} values for the sample stars are listed in cols. 8 and 9 of Table~\ref{tab_stars}, while in 
Fig.~\ref{Fig_fe_te} we show [Fe/H] as a function of the effective temperature. No notable trends of [Fe/H] as a 
function of $T_{\rm eff}$ are present. 
%, with the exception of star No.9 of NGC~6192, that is the metal richest star of the cluster 
%and also the hottest one. 
The average metallicities for the three clusters are listed in Table~\ref{tab_stars}. They are all oversolar:
[Fe/H]= $+0.12\pm0.04$ (NGC~6192), $+0.11\pm0.04$ (NGC 6404), $+0.37\pm0.03$ (NGC 6583).

As our sample is made of OCs from the inner Galaxy, they 
are expected to be metal richer than the solar neighborhood, so that
the estimated metallicities are in qualitative agreement with their
Galactocentric distances.

\subsection{Other elements} 

The [X/Fe] ratios for Mg, Al, Si, Ca, Ti ($\alpha$-elements), Cr and Ni (iron-peak elements) 
are listed in Table \ref{tab_abu}. The number of lines available for each element in each star 
is shown in parenthesis. 
Due to the very small number of lines available for Mg and Al, we computed  their ratios over iron line by line, 
and then we calculated the average. 
Errors on [X/Fe] values are
the quadratic sum of the error $\sigma$1 for  [Fe/H] and [X/H] values.
 
Mean abundance ratios [X/Fe] together with 1-$\sigma$ standard deviations are listed at the end of each 
group of stars in  Table \ref{tab_abu}. 
No evident trends of [X/Fe] ratios with [Fe/H] are present. NGC~6404 
shows a larger scatter than the other clusters due to the lower S/N of its spectra. 
The ratios of $\alpha$-elements to Fe are close to solar in the sample clusters, with the exception of [Ca/Fe] in
NGC~6192 that is slightly lower than the solar value,  $\sim-0.2$. 
Also the Fe-peak elements are solar, with [Cr/Fe] slightly below zero for NGC~6404 ($\sim$-0.2). 
NGC 6404  shows a larger scatter than  the other clusters due
to the lower S/N of its spectra.

\begin{table*}
\caption{[X/Fe] abundances and averages. Errors are the quadratic sum of $\sigma$1 on [X/H] and on [Fe/H]. }
\begin{center}
\begin{tabular}{ccccccccc}
\hline
\hline
Star 			&[MgI/Fe]&  [AlI/Fe] & [SiI/Fe] & [CaI/Fe] & [TiI/Fe] &  [NiI/Fe] & [CrI/Fe]\\
\hline
NGC6192      	&		   &				&               				& 											  &&& \\
\hline
9			&-0.16$\pm$0.16(2)& -0.08$\pm$0.14(2)&0.14$\pm$0.14(8) &-0.08$\pm$0.13(11)& -0.09$\pm$0.14(21)&0.02$\pm$0.13(34) & -0.09$\pm$0.14(10)\\		
45			&-0.01$\pm$0.16(2)&0.14$\pm$0.14(2) &0.00$\pm$0.12(6)	&    0.02$\pm$0.12(11)&	0.03$\pm$0.12(25)&-0.11$\pm$0.13(40) &0.04$\pm$0.11(12)\\	
96			&-0.07$\pm$0.11(2)&0.03$\pm$0.2(2) &	0.00$\pm$0.15(6) &	-0.04$\pm$0.15(6)	&  -0.02$\pm$0.13	(17)&-0.05$\pm$0.13(34) &-0.03$\pm$0.14(10)\\	
137		&-0.04$\pm$0.11(2)&0.03$\pm$0.08(2) &0.05$\pm$0.14(6) &	-0.09$\pm$0.13(5)	&  -0.04$\pm$0.12	(16)&-0.05$\pm$0.15(27) &0.01$\pm$0.11(9)\\
\smallskip	\\
\hline		
Average &-0.07$\pm$0.05 & 0.03$\pm$0.05 & 0.05$\pm$0.05&-0.05$\pm$0.04  &	-0.03$\pm$0.03&-0.05$\pm$0.03&	-0.02$\pm$0.04\\
\hline
\smallskip \\
NGC6404		&               				& 											  &&& \\
\hline
5 	&	0.10$\pm$0.10(1) &  0.05$\pm$0.20(2)  & 0.23$\pm$0.15(5)		&-0.24$\pm$0.15(3)	&0.08$\pm$0.13(12)	&-0.05$\pm$0.13(16) &0.20$\pm$0.14(4)\\
16 &	-0.04$\pm$0.16(2) & -0.03$\pm$0.14(1) &-0.07$\pm$0.14(5)		&-0.18$\pm$0.13(3)	&-0.15$\pm$0.14(11)	&0.13$\pm$0.13(10) &-0.02$\pm$0.16(8)\\
27 	&	-0.05$\pm$0.16(2) & 0.15$\pm$0.14(1) & 0.14$\pm$0.12(3) 		&-0.26$\pm$0.12(4)	&-0.07$\pm$0.12(11)	&0.17$\pm$0.13(14) &-0.02$\pm$0.11(8)\\
40 	&	 0.00$\pm$0.11(2) & -0.05$\pm$0.08(2) & 0.08$\pm$ 0.14(5)		&-0.13$\pm$0.13(2)	&0.12$\pm$0.12(7)	&0.00$\pm$0.15(29) &-0.04$\pm$0.14(7) \\
\smallskip \\
\hline
Average &0.00$\pm$0.05 & 0.03$\pm$0.07 & 0.09$\pm$0.09&-0.20$\pm$0.04  &	-0.01$\pm$0.10&0.06$\pm$0.08&	0.03$\pm$0.08\\
\hline
\smallskip \\
NGC6583		&               				& 											  &&& \\
\hline
46 &	-0.08$\pm$0.16(2)		&0.09$\pm$0.14(2)	&-0.05$\pm$0.14(3)		&0.00$\pm$0.13(7)	&0.01$\pm$0.14(9)	&0.02$\pm$0.13(28) &-0.16$\pm$0.15(4)\\
62 	&	-0.02$\pm$0.11(2)		&0.14$\pm$0.20(2)	&0.06$\pm$0.15(3) 		&-0.03$\pm$0.15(7)	&-0.03$\pm$0.13(12)	&0.10$\pm$0.13(22) &-0.15$\pm$0.14(7)\\
\smallskip \\
\hline
Average &-0.05$\pm$0.03&0.11$\pm$0.02 &0.01$\pm$0.05&-0.01$\pm$0.01  &	-0.01$\pm$0.02&0.06$\pm$0.04&	-0.15$\pm$0.01\\
\hline\hline
\end{tabular}
\label{tab_abu}
\end{center}
\end{table*}

\begin{figure*}[!t]
\centering
\includegraphics[angle=-90,width=1.0\textwidth]{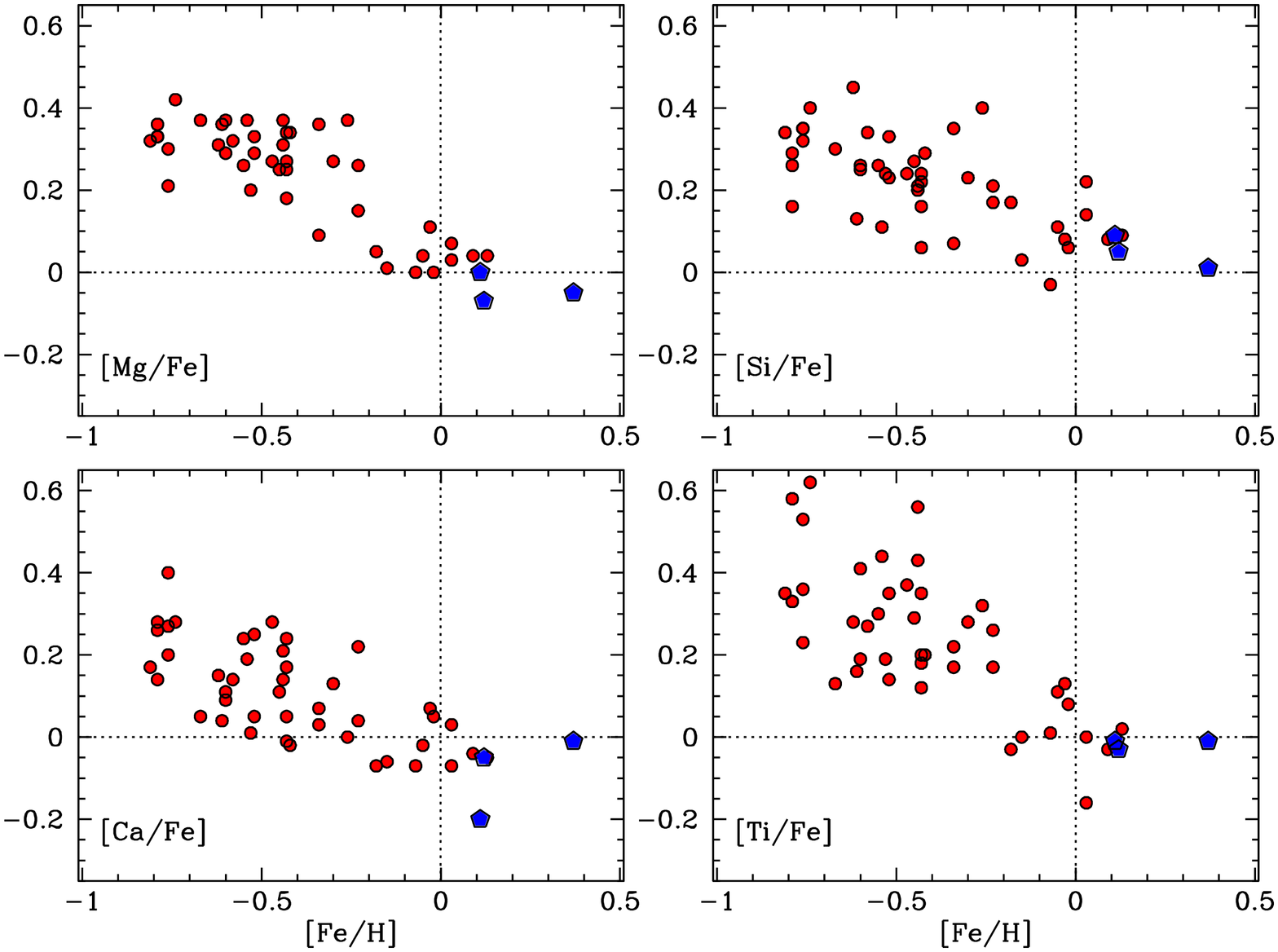} 
\caption{ The $\alpha$- elements Mg,  Ca, Si, Ti  over Fe
ratios of our  target clusters (big blue filled pentagon)  with the same  ratios recently measured
by Bensby et al. (\cite{bensby10}) for  inner disk giants (small red filled circles). }
\label{Fig_giant}
\end{figure*}

In Fig.~\ref{Fig_giant}  we  compare the $\alpha$-elements Mg,  Ca, Si, Ti  over Fe
ratios of our  target clusters with the same  ratios recently measured
by Bensby et al. (\cite{bensby10}) for  inner disk giants. The latter are located
at  Galactocentric distances  of 4-7~ kpc, hence  consistent  with the
range spanned by our targets.  The three clusters represent the metal
rich  extension of the  mean trend  of field  giants, however  they do
follow the  trend quite  smoothly. Only the  most metal  rich cluster,
NGC~6583,  might be  perhaps a  bit over-enhanced  in Ca  and  Ti.  We
interpret this behaviour as an  evidence that clusters and field stars
belong to  the same  population, i.e., they  have formed in  the same,
though extended, star formation episode.  Note that the field stars in
Bensby et  al. (\cite{bensby10})  are on average  more evolved  (mean $\log~g=1.5$)
than our cluster stars, therefore presumably slightly older, which is
consistent with the higher metallicity of the (younger) cluster stars.

\section{The metallicity gradient of open clusters}
\label{sec_grad}

The radial metallicity gradient preserves the signature of the processes of galaxy formation and evolution. 
OCs are among the best Galactic stellar populations  to study its shape and are unique tools to investigate
its time evolution. 
For this purpose, a large sample of homogeneous data in terms of metallicity, age, and distance, is necessary. 
In this respect, our starting point is  the database of OCs by M09, which is a collection of 
elemental abundances, ages and distances  derived by several authors with high-resolution spectroscopy. 
In the database, the $R_{\rm GC}$ were taken from Friel et al. (\cite{friel95}, \cite{friel02}, \cite{friel06}) or  
calculated using the cluster distance given in the original papers and  a solar Galactocentric 
distance of 8.5 kpc, consistent with Friel et al. works. 
The ages for the clusters older than 0.5-0.6~Gyr were calculated in an homogeneous way using  the morphological age 
indicator $\delta V$ (Phelps et al.~\cite{phelps94}) and the metallicity-dependent calibration of Salaris et al.~(\cite{salaris04}). 
For the younger clusters, M09 used the most recent age determinations available in the literature, such as those obtained with the lithium depletion boundary method. 
We updated the database of M09 with the latest high-resolution spectroscopic observations and with the results 
of the present work. The ages of the new clusters (only those older than $\sim$0.6~Gyr) have been recomputed homogeneously using the same method adopted by M09. For most of the sample clusters $\delta$V
is available from the literature; when not available, we obtained
this quantity from available color-magnitude diagrams. 

From the literature, we included 7 young clusters, namely  NGC~6281	(0.3~Gyr), NGC~3532 (0.35~Gyr), IC2714 (age$\sim$0.4~Gyr), NGC~2099 (0.4~Gyr), NGC~6633 (0.45~Gyr),  NGC~1883 (0.65~Gyr), 
IC~4756 (0.79~Gyr), and 7 intermediate-age clusters NGC~5822 (1.16~Gyr),  NGC~1817 (1.12~Gyr), 
Cr~110 (1.0~Gyr),   Tombaugh~2 (2.13~Gyr),
NGC~6939 (2.05~Gyr),  NGC~2158 (1.9~Gyr), and NGC~2420 (2.2~Gyr), and one old cluster Be~39 (7~Gyr). 
Abundance determinations are from Smiljanic et al.~(\cite{sm09}) (IC~2714, IC~4756, NGC~3532, NGC~6281, NGC~6633),  Jacobson et al. (\cite{ja09}) (NGC~1817, NGC~1883), Villanova et al.~(\cite{villanova10}) (To~2), Pace et al. (\cite{pace10}) (NGC~5882),  Pancino et al. (\cite{pancino10}) (Cr~110, NGC~2099, NGC~2420), Friel et al.~(\cite{friel10} (Be~39).
We also updated the M09 database with the results by Friel et al.~ (\cite{friel10}) for Be~31. 
NGC~6404 and NGC~6192 are too young to be 
calibrated with the Salaris et al.'s relationship, while the  age of NGC~6583, the oldest cluster of our sample, 
computed with the morphological age indicator $\delta$V is $\sim$1~Gyr, in agreement with the age given by 
Carraro et al.~(\cite{carraro05}) with isochrone fitting. 

\subsection{The inner gradient}

\begin{figure*}[!t]
\centering
\includegraphics[angle=-90,width=1.\textwidth]{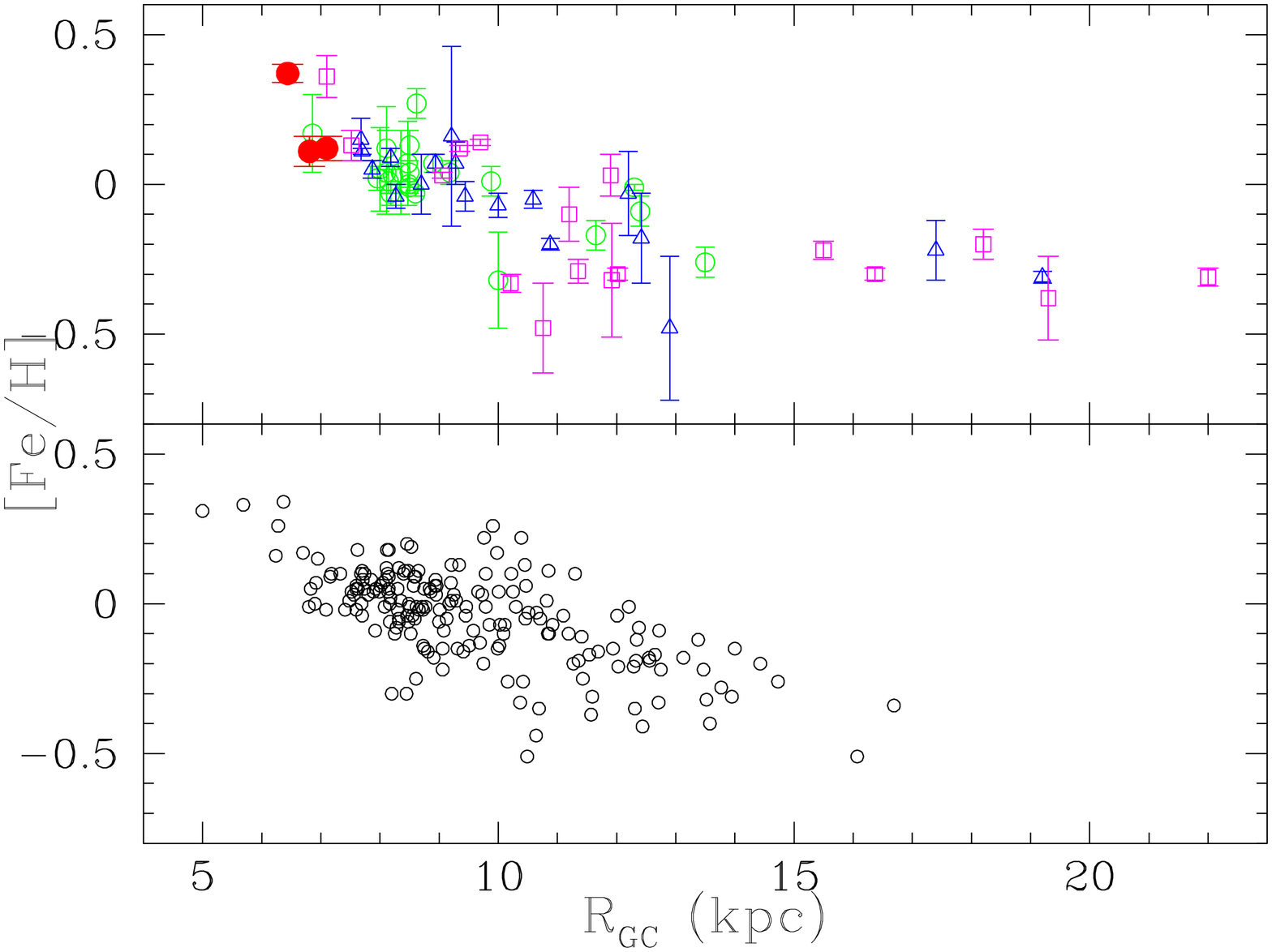} 
\caption{The iron gradient: OCs of different ages (top panel) and Cepheids (bottom panel).
Empty  circles (green) are OCs with age$\leq$0.8~Gyr, triangles (blue) with 0.8~Gyr$<$age$\leq$4~Gyr, and squares (magenta) with age$>$4~Gyr. 
The three large filled circles (red) are 
the data presented in this paper. The black small empty circles  in the lower panel are [Fe/H] of Cepheids by Pedicelli et al. (\cite{pedicelli09}). 
%The continuous magenta
%lines are the least mean square fits to the Cepheid data in the radial range 5-8~kpc and 8-16~kpc.  
%The three curves are the model by M09 at present time (continuous green), 2 Gyr ago (blue dashed) and 4 Gyr ago
%(cyan dotted).
}
\label{Fig_grad_inner}
\end{figure*}

The behaviour of the metallicity gradient in the inner part of the Galactic disk (R$_{\rm{GC}}<$8~kpc)
is still an open issue. The inner part of the disk is of paramount importance because it connects 
the chemistry of the disk with that of the bulge. 

Several studies identified one or more changes of slope in the [Fe/H] radial distribution.  
For OCs, Twarog et al. (\cite{twarog97}) firstly  identified a break in the [Fe/H] radial distribution of their cluster sample at
R$_{\rm{GC}}\sim$10 kpc and this was recognized as the first indication of a transition between the inner and 
outer Milky Way disks (cf. Friel et al.~\cite{friel10}). 
The occurrence of a steepening of the metallicity gradient in the inner disk with respect to the outer regions 
was also suggested  by Andrievsky et al. (\cite{an02}) who found, using Cepheid abundances, 
two changes of slope in the radial gradient
at  $R_{\rm GC} \sim 11-12$ kpc and $R_{\rm GC} \sim 7$ kpc. 
The large compilation of Cepheid data by Pedicelli et al. (\cite{pedicelli09}) shows instead 
a smooth and steady increase in the slope for $R_{\rm GC} < 8$ kpc, without any further steepening towards the GC. 
They found an  iron gradient in the inner disk ($R_{\rm GC} < 8$ kpc) more than a factor of three steeper ($-0.130\pm0.015$ dex kpc$^{-1}$) than in the outer disk  ($-0.042 \pm 0.004$ dex kpc$^{-1}$). 
Neither the metallicity gradients of planetary nebulae (PNe),  nor that  of HII regions show any evident steepening in the inner regions 
(cf. Stanghellini \& Haywood \cite{stanghellini10}, Deharveng~\cite{deha00}). In the case of PNe and HII regions,  gradients of 
oxygen and other $\alpha$-element (i.e. S/H, Ne/H, Ar/H) were derived, rather than for iron; 
thus a direct comparison is more difficult.  In addition, 
few of these objects have been studied at $R_{\rm GC}>15$~kpc.

In Fig.~\ref{Fig_grad_inner} we compare the iron gradient of OCs with that of Cepheids (data from Pedicelli et al. \cite{pedicelli09}) 
in the radial range 4~kpc$<R_{\rm GC} < $23~kpc. 
%The OCs are marked with different symbols accordingly to their age: filled small circles are OCs with age$\leq$0.8~Gyr, 
 %triangles with 0.8~Gyr$<$age$\leq$4~Gyr, and squares with age$>$4~Gyr. The three large filled circles are 
%the data presented in this paper. The empty circles are [Fe/H] of Cepheids by Pedicelli et al. (\cite{pedicelli09}). 
First of all, the metallicity scales of the two populations are in very good agreement. 
The gradients of young OCs  and of Cepheids compare  very well in the whole radial range. A weighted mean least-square fit 
of the whole sample of young OCs (age$\leq$0.8~Gyr) 
gives for $R_{\rm GC} \leq 8$ kpc a slope of $-0.137\pm0.041$ dex kpc$^{-1}$ and for $R_{\rm GC} > 8$ kpc$-0.033 \pm 0.005$ dex kpc$^{-1}$, 
which are in agreement with the values $-0.130\pm0.015$ dex kpc$^{-1}$ and $-0.042 \pm 0.004$ dex kpc$^{-1}$ of Pedicelli et al.~(\cite{pedicelli09}). 
%Then, both the average metallicity and the metallicity gradient of the intermediate-age and old clusters  do not differ very much from 
%the young population one (young clusters and Cepheids) indicating a little evolution of the Galactic disk metallicity during the last Gyrs  (see M09 for a complete discussion). 
The main differences between young (Cepheids and young OCs) and old (intermediate-age and old OCs) stellar populations reside in the outer disk, where 
several old and intermediate-age clusters are observed. 
Their radial metallicity distribution becomes flat at large Galactocentric radii, while this effect is not appreciable
in the young populations due to the lack of measurements.  

Both populations show a clear steepening of the gradient at 
similar values of R$_{GC}$. However, due to the high dispersion at each R$_{GC}$, 
it is difficult to identify the exact radius where the change of slope happens.  
%In addition, the increase of metallicity in the inner-disk  Galaxy as traced by young OCs (age$<$1~Gyr) does not reach 
%extremely oversolar values, but stops when it reaches the maximum value, defined by the stellar iron yield (cf., e.g., Pilyugin et al.~\cite{py07} for the maximum stellar  yield of oxygen).
%The presence of a change of slope in the inner disk metallicity gradient has 
%important consequences for our understanding
%of the processes controlling the Galaxy  formation and evolution. 

As described in the Introduction, several chemical evolution models can 
reproduce changes of slope in the abundance gradient (e.g., Lepine et al.~\cite{lepine01}, M09, Colavitti et al.~\cite{colavitti09}, 
Fu et al.~\cite{fu09}) making different assumptions and hypothesis. 
Without entering in model details, the observational evidence of a steeper inner gradient
suggests  that the inner disk has evolved at a different rate with respect to the outer disk. 
Many reasons can be at the origin of this; among others we 
recall the predominance of the halo collapse in the inner disk,  the effect of the corotation
resonance, the existence of density threshold in  star formation, and/or radial varying star formation efficiency. 
Combinations of these causes are also likely to operate 
together.
However,   before a definitive interpretation in terms of models could be given, data still demand some further investigations
and analysis. For example, there is still very little information about the possible azimuthal variations of the gradient 
(e.g., Luck et al. \cite{luck06}; Stanghellini \& Haywood, \cite{stanghellini10}).

\subsection{The time-evolution of the inner gradient}

\begin{figure}[!t]
\centering
\includegraphics[width=0.45\textwidth]{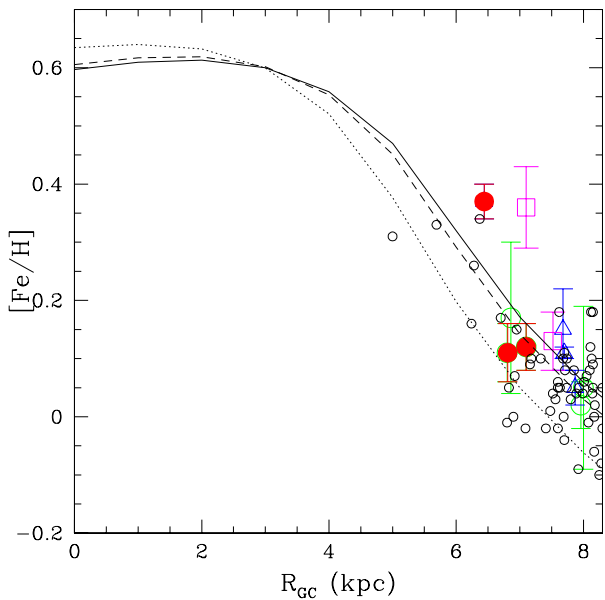} 
\caption{The disk inner gradient  (0~kpc $< R_{\rm GC} < 8$~kpc).  Symbols for OCs and Cepheids 
 are as in Fig.~\ref{Fig_grad_inner}. We show also the model curves of M09 at different epochs: present-time 
 (continuous line),  2~Gyr ago (dashed curve), 6~Gyr ago (dotted curve).   }
\label{Fig_grad_inner_ev}
\end{figure}

In Fig.~\ref{Fig_grad_inner_ev} we show the iron gradient in the very inner regions, between 0 and 8~kpc 
form the GC. For $R_{\rm GC} < 4$~kpc OC and Cepheid metallicities are not available, and 
we do not show metallicity measurements 
for the disk populations because of the difficulty to disentangle between disk and bulge stars. 
We compare the inner gradient traced by  young populations, such as  Cepheids and young clusters, with 
the gradient of older OCs. 
%The comparison is done visually because of the paucity of young OC measurements 
%for $R_{\rm GC} < 8$~kpc respect to those of Cepheids (5 vs. 50). 
The [Fe/H] gradient of young clusters  agrees with the Cepheid one, as expected 
for populations with similar ages and both NGC~6192 and NGC~6404 are in perfect agreement
with respect to the average abundance of Cepheids.
 
On the other hand, the [Fe/H] values of some old and intermediate-age OCs are surprisingly higher 
than for  younger clusters; this is true  in particular for the two innermost intermediate-age/old OCs: NGC~6583 (from this paper) 
and NGC~6253 (Sestito et al. \cite{sestito07}).
The predictions of 'classical' evolution models for the Galactic disk, assuming that the disk
itself is formed by the collapse of the halo (with one or more processes of infall) and radial mixing is not considered, are 
in this respect well known. They imply that at each radius the metallicity increases with time, 
so that the older is the population considered, the lower is the metallicity 

This is  shown in Fig.~\ref{Fig_grad_inner_ev} in the case of the chemical evolution model by M09 
(see also, e.g., several cases of the two-infall model shown in Fig.~11 by Chiappini et al.~\cite{chiappini01}). 
The M09's model reproduces only the disk evolution and does not include the bulge.  
The three curves represent the iron gradient predicted by the model  at three different epochs in the Galaxy lifetime 
(present-time, 2~Gyr ago, 6~Gyr ago). 
%From the Galaxy centre to R$_{GC}\sim$4~kpc the 
%gradient is almost flat because the iron abundance has  reached its maximum value, which 
%is defined by the stellar iron yield, i.e. the mass of iron freshly synthesized and 
%ejected by a generation of stars relative to the mass locked up in low-mass stars and compact remnants.
%It can be derived within the framework of chemical evolution models of galaxies (Pagel 1997). The closed-box model which neglects mass exchange between a galaxy and its environments gives the maximal upper bound to the metallicity of the gas for a given gas mass fraction (Edmunds 1990). 
For R$_{\rm{GC}}\geq$4~kpc the iron abundance decreases as  the Galactocentric radius increases; 
at a given radius the metallicity at early epochs is lower than at present.  
Only in the very inner region, $R_{\rm GC} < 3$~kpc, the gradient at early epochs and 
the gradient at present time are inverted, i.e. old population are slightly metal richer than  young ones. 
This is due to the almost complete conversion of the gas into stars already at early epochs, 
and to the subsequent dilution at recent times with gas expelled during the final phases of low-mass star evolution.

One could imagine to change the model parameters in order to move the Galactocentric radius of the point of inversion of the metallicity 
gradient, and to reproduce the high metallicity of some of the old and intermediate-age clusters in the inner Galaxy.  
The model parameter that governs the position of the crossing-point is primarily the scale-length of the infall law 
(see M09 for detail). Varying the scale-length of the exponential law driving the infall of gas from the halo, 
we can produce a more or less extended inner metallicity plateau. However reasonable values 
of the scale-length can vary the location only from $\sim$2 to $\sim$4 kpc form the GC, and 
cannot reproduce the position, in the [Fe/H] versus R$_{\rm{GC}}$ plot, of clusters like NGC~6583 and NGC~6253.

The location of these two clusters needs different explanations, such as, 
a different  place of birth with respect to their present-time location or  an incorrect  (over-estimated) distance estimates. 
Non-circular orbits  and/or  radial migration of populations (stars, but also gas) can
explain the large scatter of [Fe/H] at each radius and at each age in the Galactic disk 
(see, e.g., the chemical evolution model of Sch{\"o}nrich 
\& Binney~2009). In fact, not only clusters, but also field stars could have significantly non-circular orbits
orbits, with apocentric and pericentric distances 
differing by more than $\sim$1~kpc.
To verify this hypothesis, we have analyzed the orbits of the complete sample of inner disk OCs.

\section{The orbits of the inner OCs}
\label{sec_orbit}

In this Section we describe the results of orbit calculations for a sample of inner-disk clusters.
Firstly, we illustrate how absolute proper motion components are derived, and then we calculate
clusters' orbits adopting a variety of Galactic potential models. The main aim is to investigate
whether, for clusters located inside the solar ring, the bar can significantly affect their orbit
or not. For this result, among the various models, we will focus only on the bar-less Allen \& Santillan (\cite{allen91})
model and the Ferrers $n=2$ model (Binney \& Tremaine~\cite{binney08}), which has the maximum bar strength.
Finally, we will use the orbital parameters to address the effect of the orbital motion on the chemical
evolution of these clusters. Such an exercise is a classical one. Examples of that are in Carraro \& Chiosi (\cite{carraro94}),
while open cluster orbits have been calculated several times in the past (see e.g. Bellini et al. \cite{bellini10}
or Carraro et al.~\cite{carraro06} for recent examples). However, in all past cases, 
no effect of the bar could be seen, due to the mean large distance of the clusters from the GC and
the relatively small sphere of influence of the bar.
For the clusters under analysis here, no previous attempts have been done to derive their orbits.

\subsection{Proper motion components}

In order to obtain the clusters' mean absolute proper motion components and membership probabilities 
we took proper motion data (and associated uncertainties) on individual 
stars from the UCAC3 catalogue (Zacharias et al. \cite{zac10}). Here
we briefly recall the basics of the method. The interested reader can find all the details
in Jilkova et al. (in preparation). 

The basic method to segregate cluster and field stars using proper motion distributions was 
described long time ago by Sanders (\cite{sanders71}). 
The fundamental 
assumption of this model is that proper motions are distributed according to circular or elliptical bi-variate normal distributions
for  cluster and for field stars, respectively. A maximum likelihood principle is then used to determine the distributions' parameters. 
Zhao \& He (\cite{zh90}) later on  improved the method by weighting proper motion components  with observational uncertainties. 

We devised the 9 parametric models starting from the description by Wu et al. (\cite{wu02}), but keeping the dispersion of the circular Gaussian distribution of 
clusters stars as zero.
Therefore the width of the distribution is given only by the observed errors of proper motions (see also the discussion in 
Balaguer-N{\'u}{\~n}ez  et al. \cite{bn04}).

From the UCAC3 catalogue we selected only those stars having measurements of their proper motions and  errors,  and employed the 2MASS 
photometry as an additional check. The choice of the sample radius around each cluster center was made individually for each cluster 
taking into account the suggestions by Sanchez et al. (\cite{sanchez10}).
To minimize the influence of possible outliers, we firstly considered all the stars 
belonging to one population with a single bivariate normal distribution. Further we used only stars with both components 
of proper motion within the 3-$\sigma$ interval. Then we restricted the range of the sample to $|\mu| < 30$ mas/yr.
For such a sample the two-distributions analysis was done. The solution of maximum likelihood equations was found
by a non-linear least squares minimization procedure. 

Once the membership probabilities from pure proper motion data were obtained, the CMDs using the 2MASS J magnitudes vs. 
(J-H) color indexes were inspected for the stars with cluster membership probability higher than 90$\%$. For some clusters
the additional magnitude constraint was used to eliminate field stars and iteratively the parameters of the new 
distribution were obtained. For more details and a comparison with previously derived UCAC2 proper motions,  see Jilkova et al. (in preparation).
In Table 10 we give the identification name  of the clusters, their equatorial coordinates (RA, Dec), their Galactic coordinates (l and b), their distances from the Sun, 
their average radial velocities, proper motion in mas/yr ($\mu_{\alpha}$ and $\mu_{\delta}$),  locations in the Galactic system (X, Y, Z) ,  radial velocity components (U, V, W), and finally age and metallicity [Fe/H].

\subsection{Orbit computation}

With absolute proper motions, radial velocities and heliocentric distances of clusters we calculated their
Galactic orbits. We assumed the heliocentric distances with 10\% errors and we derived three sets of initial 
conditions for each cluster. The initial conditions were calculated following the procedure by Johnson \& Soderblom (\cite{js87}).
We adopted the solar motion components with respect to LSR as being (U,V,W) = (11.1,12.4,7.25) km/s from 
Sch{\"o}nrich et al. (\cite{sc10}).
We also used a
right-handed
coordinate system where $U$ is positive in the direction to the GC, the rotation velocity of the 
LSR of 220\,km$/$s, and the Galactocentric distance of the Sun is 8.5\,kpc. 

For the Galactic potential we adopted the axisymmetric time-independent model by Allen \& Santillan (\cite{allen91}) and we 
also considered an influence of the Galactic bar. For the latter we used a Ferrers potential ($n=2$) of 
inhomogeneous triaxial ellipsoids (Pfenniger~\cite{pf84}). The characteristics of the bar model are: 
{\em i)} the bar 
length of 3.14 kpc; {\em ii)} axial ratios 10:3.75:2.56; {\em iii)} a mass of $9.0\cdot 10^{10}$\,M$_{\odot}$;
{\em iv)} an
angular speed of 60 km/s/kpc: and {\em v)} an initial angle with respect to Sun--GC direction of 
20$^{\circ}$. We used the same parameters values as Pichardo et al. (\cite{pic04})  or Allen et al. (\cite{allen08}) for their more
sophisticated models of the Galactic bar. For the axisymmetric background potential we kept the Allen \& Santillan (\cite{allen91}) 
model,  except for  the mass of the bulge of $4.26\times 10^{9}$\,M$_{odot}$ (the bar replaces 70\% of bulge mass).

The integration routine uses a Runge-Kutta 4th order integrator (Press et al.~\cite{press92}). The relative change in the 
Jacobi constant is of the order of $10^{-9}$ to $10^{-10}$. We integrated the orbits backwards in time over
the intervals corresponding to the age of the clusters. For each orbit we give the most relevant
orbital parameters (mean peri- and apo-galacticon, maximum height from the plane, eccentricity, and birthplace) 
in Table~\ref{tab_orbit}. Errors associated to each of these values are computed averaging the values obtained
by running orbit calculations with 3 different distances.

\begin{figure}
 \centering
  \includegraphics[angle=0,width=9cm]{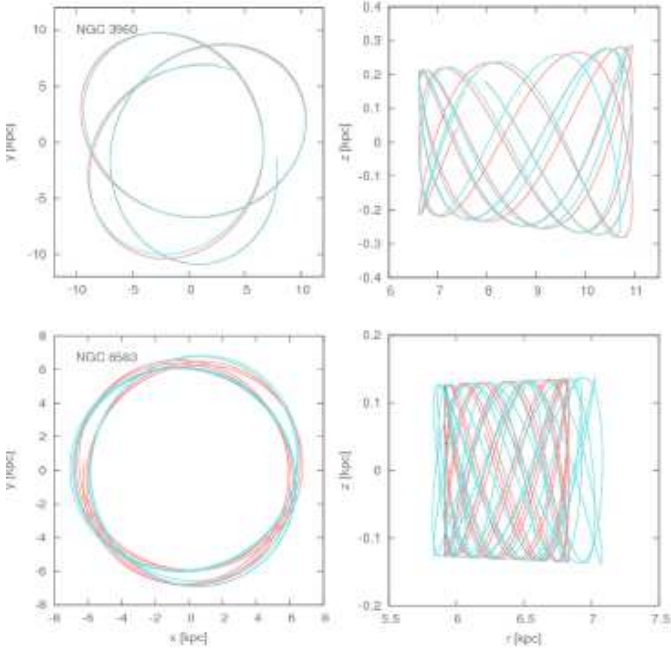}
   \caption{Orbit computation for two clusters, NGC~3690 and NGC~6583, for the bar-less Allen \& Santillan (1991) model
(red), and the Ferrers $n=2$ model (green).}
\label{Fig_task}%
\end{figure}

\noindent
In Fig.\ref{Fig_task} we show, as an illustration, the orbit computation for two clusters, NGC~3690 and NGC~6583, for the bar-less Allen \& Santillan (1991) model
(red), and the Ferrers $n=2$ model (green), to show a case where the bar does not
have any influence (NGC~3690) and a case where the bar  slightly affects  the orbit  (NGC~6583).

\begin{figure}
  \centering
   \includegraphics[angle=0,width=8cm, height=8cm]{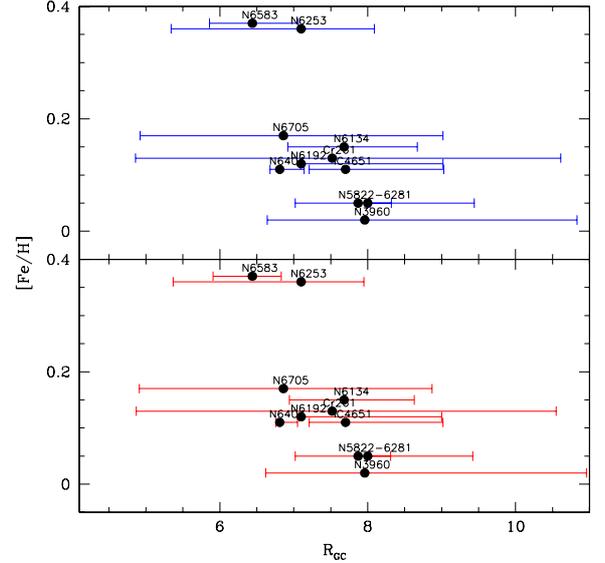}
    \caption{The epicyclic variation of the radial position of the considered OCs. In this plot we show the results from orbit calculations  for the bar-less Allen \& Santillan model (upper panel) 
    and for the bar Ferrers model (lower panel).  }
\label{Fig_epi}%
\end{figure}

In Fig.\ref{Fig_epi} we show the effect of the OC orbital motion on their radial position. For each cluster an horizontal bar shows the epicyclical variation 
from the apogalacticon to the perigalacticon. For some clusters, with eccentric orbits, large variations of their radial positions, up to about 4 kpc,  are allowed during their lifetime. 

\subsection{The influence of orbits on the shape of the gradient}
\begin{figure*}
  \centering
   \includegraphics[angle=0,width=15cm]{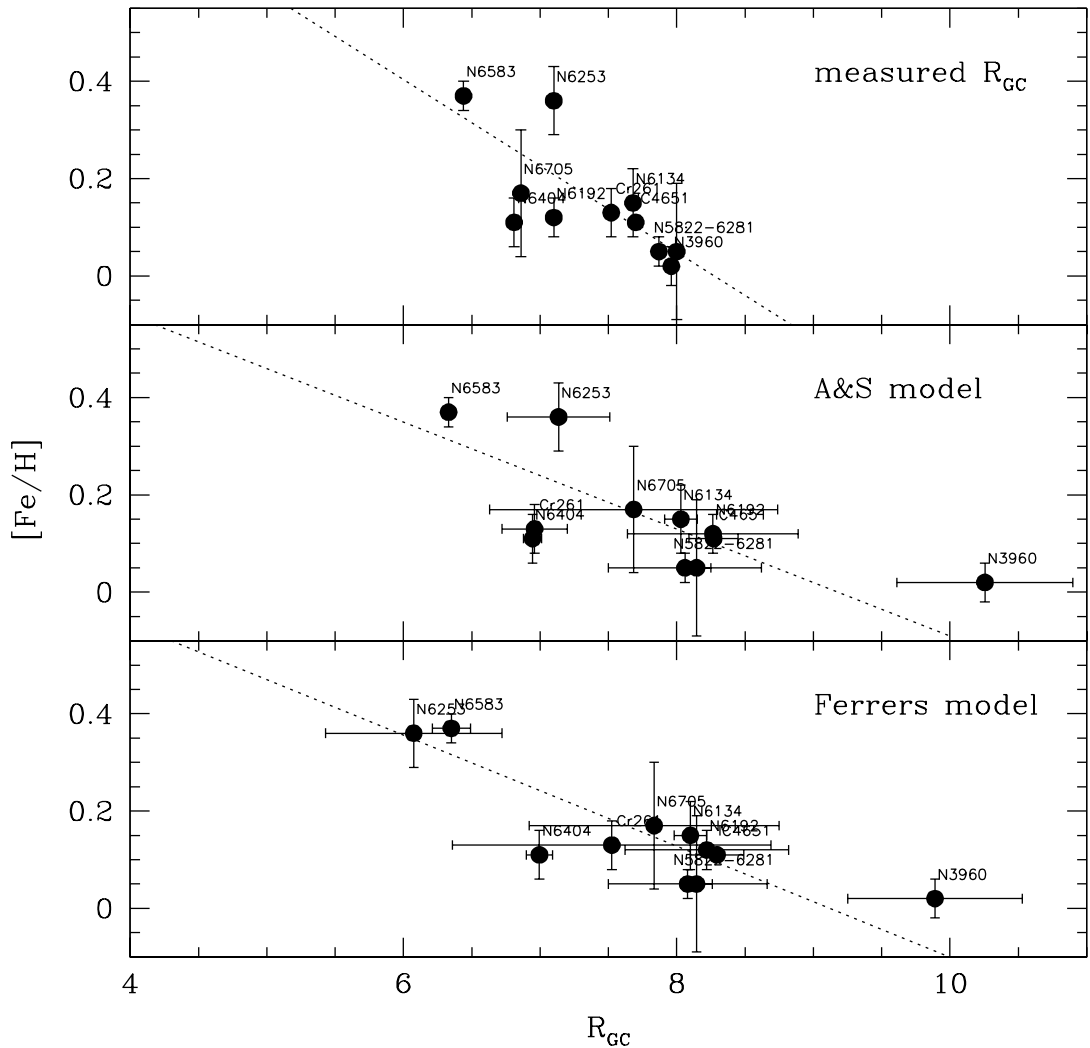}
    \caption{The inner metallicity gradient of OCs with: 
    measured Galactocentric distances (upper panel), radial position at the time of birth of the clusters, from orbit calculations  for the bar-less Allen \& Santillan model (middle panel), 
    and from orbit calculations for the n=2 Ferrer bar (lower panel).  The dotted lines are average least-squared fits.  }
\label{Fig_orbit}%
\end{figure*}

The most interesting quantity in order to derive the original shape of the metallicity gradient is the cluster birthplace. 
We have computed it integrating the orbits up to the epoch of birth of each cluster (see Table~\ref{tab_par1}). The birthplaces, 
computed  with the maximum age  (Age $+$ $\Delta$Age) and with the 
minimum age (Age $-$ $\Delta$Age)  are shown 
in Table~\ref{tab_orbit}. We have assumed that $\Delta$Age is $\sim$10\%.
The uncertainty  due to the error on the cluster distance is estimated integrating the cluster's orbit with three different 
distances, R$_{\rm{GC}}$,   R$_{\rm{GC}} \pm \Delta R_{\rm{GC}}$, and it is represented by the errors on birthplaces 
in  Table~\ref{tab_orbit}.
%As shown in Fig.\ref{Fig_orbit}, for the younger clusters of our sample (IC~4651, NGC~3960, NGC~5822, NGC~6192, NGC~6281, 
%NGC~6404, NGC~6583, NGC~6705 with age$\leq$ 1 ~Gyr) 
%the  uncertainty  on their age  has a small effect on their birthplace.  
%On the other hand, in the case of the oldest clusters of our sample (Cr261--8.4~Gyr and NGC~6253--4.5~Gyr) 
%the uncertainty  on their age (of the order of $\sim$1~Gyr)
%is the main cause of error on the determination their place of birth, as shown in Fig.\ref{Fig_orbit}. 
%In addition, the hypothesis of constant Galactic gravitational potential with time is not well 
%verified when the integration goes backwards for $\sim$8~Gyr. 

In Fig.~\ref{Fig_orbit} we show the metallicity gradient computed with three different sets of radial positions:  {\em i)} the measured R$_{\rm{GC}}$, {\em ii)}
the radial positions at the time of birth of the clusters from orbit calculations  assuming  the bar-less Allen \& Santillan model,  
and {\em iii)} the R$_{\rm{GC}}$ birthplace from orbit calculations with the $n=2$ Ferrers bar model. 
From  Figs.~\ref{Fig_epi} and  \ref{Fig_orbit}, we  test our initial hypothesis of a different  place of birth with respect to their present-time location 
of NGC~6253 and NGC~6583 and we check how much the present-time shape of the gradient is  affected by cluster orbital motion. 
The initial radial position of some clusters differs remarkably from their present-time position. In particular, for the clusters closer to the GC 
whose orbits might be affected by the presence of a bar, the choice of a model with or without bar is  important to derive their birthplaces.    

NGC~6253 has a quite eccentric orbit and its radial position ranges from the perigalacticon  $\sim$5.4~kpc to the apogalacticon 
$\sim$8~kpc, both with bar and bar-less models (see also Fig.~\ref{Fig_task}).  
With an age estimate of $\sim$4.5$\pm$0.5~Gyr, the birthplace of this cluster would be $\sim$6~kpc with the bar model or $\sim$7~kpc 
with the bar-less model, hence moved inward with respect to its present-time location.  
However,  the age of NGC~6253 is particularly uncertain:  Piatti et al. (\cite{piatti98}) found an age of 5$\pm$1.5 Gyr,   Twarog et al. (\cite{twarog03}) derived 3$\pm$0.5 Gyr, 
and Bragaglia \& Tosi (\cite{bra06}) indicated an age of $\sim$3~Gyr. 
The uncertainty on the age impacts on the position of the cluster along the orbit and might allow even more internal birthplace, up to $\sim$5.4~kpc (the perigalacticon), 
as expected from its high metallicity. \\
 NGC~6583 is younger than NGC~6253, with an age of $\sim$1~Gyr, thus the propagated uncertainty due to age determinations on its birthplace is smaller. 
In addition, the orbit of NGC~6583 is quite circular, with an epicyclic variation of $\sim$1~kpc.
With both Galactic potential models, the birthplace of NGC~6583 is expected to have a small variation with respect to its present-time 
location. Thus NGC~6583 is really a high-metallicity cluster located at about R$_{\rm{GC}}=$6.4~kpc.

Weighted linear fits of the gradient slope  are shown in Fig.~\ref{Fig_orbit}  for the 
 R$_{GC}$, for birthplaces  computed either 
with bar-less or with bar models. 
The gradient with measured R$_{GC}$ has a steep slope  driven by the position of NGC~6253. 
The three slopes   are $-$0.177$\pm$0.034 dex kpc$^{-1}$, $-$0.110$\pm$0.026 dex kpc$^{-1}$, $-$0.114$\pm$0.021 dex kpc$^{-1}$, respectively.
We excluded NGC~3690 from the fits because its birthplace is in a different radial region 
where the gradient changes its slope, becoming flatter.   
The inner gradient appears more defined  and in better agreement with Cepheids data ($-$0.130$\pm$0.015~dex~kpc$^{-1}$, cf. Pedicelli et al.~\cite{pedicelli09})
when we consider  the initial 
radial position of each cluster, in particular for the birthplaces computed with the bar model. 

Thus, concluding, the knowledge of OC orbits is an important tool to derive correctly the shape and slope of the metallicity gradient. 
 Even if statistically the effect might be marginal (see also, e.g., Wu et al. \cite{wu09}), the determination of clusters'  orbits and birthplaces 
 is important to clarify the position of some 'anomalous' clusters, such as NGC~6253, or to confirm their position, as in the case of NGC~6583.

\section{Summary and conclusions}
\label{sec_sum}

In this paper we have presented new high-resolution spectroscopic observations of evolved stars in three OCs located 
between 6 and 7 kpc from the GC (NGC~6192, NGC~6404, NGC~6583) together with orbit computations for  
the complete sample of clusters with R$_{\rm{GC}}$$<$8~kpc and available 
high-resolution spectroscopy. 
The UVES observations allow us to extend the sample of clusters studied in the inner Galactic disk and to discuss the shape of the  
metallicity gradient in this radial region.

The main results we have obtained are the following: 
\begin{itemize}
\item[i)] from radial velocity analysis, we obtained the membership of  evolved stars finding 4 members in NGC~6192, 4 in NGC~6404, and 2 in NGC~6583;
\item[ii)] from their high-resolution spectra  and standard LTE analysis,  we derived stellar parameters and abundance ratios of the iron-peak elements Fe, Ni, Cr, and of the $\alpha$-elements Al, Mg, Si, Ti, Ca. 
Their  average metallicities  are [Fe/H]= $+0.12\pm0.04$ (NGC~6192), $+0.11\pm0.04$ (NGC 6404), $+0.37\pm0.03$(NGC 6583) and  [X/Fe] are consistent with solar values. In addition, we have compared our results with abundance ratios of inner disk giant 
stars finding a good agreement;
\item[iii)] the clusters we have analysed, together with other OC and Cepheid data, confirm a steep gradient in the inner disk, 
a signature of a different evolutionary rate with respect to the outer disk.   
%Young clusters (age$<$1~Gyr) in the inner disk (R$_{\rm{GC}}$$<$8~kpc) demonstrate  that  metallicity does not increase 
%up to extremely oversolar values, but reaches a saturation due to the complete consumption of the 
%gas and to the maximum iron yield;
\item[iv)] we found that, at a given radius,  the metallicity of some intermediate-age and old clusters
in this radial region is higher than that of young clusters (in particular this is true for NGC~6583 and NGC~6253). 
For the first time, we have coupled metallicity studies with dynamics, computing the orbits of inner disk OCs and, from them, 
the birthplaces of the OCs with R$_{\rm{GC}}$$<$8~kpc. 
We have found that there is a high probability that NGC~6253 had more internal value of R$_{\rm{GC}}$ than
shown by its present position; this would reconcile its high metallicity with  its present-time radial location.  
NGC~6583 has instead a higher probability to
be born close to its present-time position, and thus to be the most internal and metal-rich cluster cluster studied; 
\item[v)] the gradients computed using the cluster birthplaces instead of their present R$_{\rm{GC}}$ distances
do not yield significantly different values for the final slopes. 
However we found a better agreement with the Cepheid inner gradient slope and a higher correlation  
when we consider the birthplaces instead of the present-time R$_{\rm{GC}}$. 
\end{itemize}

\begin{landscape}
\begin{table}
%\centering
\scriptsize
\begin{tabular}{ccccccccccccccccc}
\hline\hline
Cluster 	& RA 	& DEC	& l 		& b 		& d		& RV 		& $\mu_{\alpha}$ & $\mu_{\delta}$ & X 	& Y 	   & Z 	& U 	  & V 	    & W          & age & [Fe/H]\\
		&\multicolumn{2}{c}{J2000.0}  &&		&		&			&					&					&	&	&		&	&		&		&	&\\
\hline
		& hh:mm:ss 	& dd:mm:ss 		& deg 	& deg 	& pc 	&  km/s 	& mas/yr 			& mas/yr 				   & kpc & kpc & kpc  & km/s & km/s & km/s & Gyr &\\
\hline
Cr~261  &   12:37:57& -68:22:00&  301.684&   -5.528&  2400&     -25.43$\pm$1.11&    -5.71$\pm$1.39&     0.50$\pm$1.59   & 6.745&   -2.033&   -0.231&   -58.67&    212.38&     11.73&  8.40 &0.13$\pm$0.05 \\
IC~4651 &   17:24:49& -49:56:00&  340.088&   -7.907&   890&     -30.98$\pm$0.11&    -1.62$\pm$1.07&    -2.99$\pm$1.05    &7.173&   -0.300&   -0.122&   -23.91&    222.34&     10.10&  1.70 & 0.11$\pm$0.01\\
NGC~3960&   11:50:33& -55:40:24&  294.367&    6.183&  1680&     -22.26$\pm$0.36&    -4.16$\pm$3.58&     1.89$\pm$4.08 &   7.311&   -1.521&    0.181&   -32.06&    231.32&     11.49&  0.70 &0.02$\pm$0.04 \\
NGC~5822&   15:04:21& -54:23:48&  321.573&    3.593&   770&     -29.31$\pm$0.18&    -6.99$\pm$0.58&    -5.37$\pm$0.59   & 7.398&   -0.478&    0.048&   -32.52&    218.30&      0.61&  1.00 &0.05$\pm$0.03\\
NGC~6134&   16:27:46& -49:09:06&  334.917&   -0.198&   910&     -25.70$\pm$0.19&    -1.65$\pm$1.44&    -6.41$\pm$1.62   & 7.173&   -0.387&   -0.003&   -23.93&    213.58&     -6.92&  1.60 &0.15$\pm$0.07\\
NGC~6192&   16:40:23& -43:22:00&  340.647&    2.122&  1500&      -7.70$\pm$0.38&     0.48$\pm$1.87&     1.43$\pm$1.67    &6.586&   -0.497&    0.056&     5.82&    237.15&     10.99&  0.20 &0.12$\pm$0.04\\
NGC~6253&   16:59:05& -52:42:30&  335.460&   -6.251&  1580&     -29.71$\pm$0.79&    -4.27$\pm$3.44&    -4.88$\pm$3.20  &  6.571&   -0.652&   -0.172&   -36.75&    193.30&     13.12&  4.50 &0.36$\pm$0.07\\
NGC~6281&   17:04:41& -37:59:06&  347.731&    1.972&   480&      -5.58$\pm$0.26&    -2.30$\pm$2.11&    -4.57$\pm$1.84    &7.532&   -0.102&    0.016&     2.16&    215.26&      4.74&  0.30 &0.05$\pm$0.14\\
NGC~6404&   17:39:37& -33:14:48&  355.659&   -1.177&  1700&     -10.60$\pm$1.10&     0.75$\pm$2.80&    -2.32$\pm$3.48  &  6.305&   -0.129&   -0.035&    -1.87&    213.53&     -7.79&  0.50 &0.11$\pm$0.05\\
NGC~6583&   18:15:49& -22:08:12&    9.283&   -2.534&  2100&      -3.00$\pm$0.40&    -1.12$\pm$2.68&    -1.45$\pm$2.76    &5.930&    0.338&   -0.093&    10.04&    206.99&     10.19&  1.00 &0.37$\pm$0.03\\
NGC~6705&   18:51:05& -06:16:12&   27.307&   -2.776&  1880&      35.08$\pm$0.32&    -4.98$\pm$2.86&    -3.05$\pm$4.13   & 6.334&    0.860&   -0.091&    62.58&    202.57&     32.51&  0.20 &0.17$\pm$0.13\\
\hline\hline
\end{tabular}
\label{tab_par1}
\caption{Parameters of the OCs located within $\sim$8 kpc from the GC.}
\end{table}

\begin{table}
\scriptsize
\centering
\begin{tabular}{ccccccc}
\hline\hline
Cluster & Perigalacticon & Apogalacticon & $Z_{mean}$ & e & Birthplace (age-$\Delta$age) & Birthplace (age+$\Delta$age) \\
\hline
& kpc & kpc & kpc & & kpc\\
\hline
 Cr~261   &  	     4.87$\pm$0.23 &   10.55$\pm$0.04 &    0.161$\pm$0.011  &   0.369$\pm$0.022 &7.20$\pm$0.63 &      6.72$\pm$0.59\\
 IC~4651  &	     7.21$\pm$0.08 &    9.02$\pm$0.08 &    0.168$\pm$0.004  &   0.112$\pm$0.001 &8.45 $\pm$0.48  &     8.09 $\pm$0.59\\
 NGC~3960 &	     6.62$\pm$0.14 &   10.96$\pm$0.03 &    0.248$\pm$0.013  &   0.247$\pm$0.012 &9.61 $\pm$0.30  &    10.90 $\pm$0.03\\
 NGC~5822 &	     7.02$\pm$0.11 &    9.42$\pm$0.06 &    0.050$\pm$0.002  &   0.146$\pm$0.005 &8.62 $\pm$0.50  &     7.50 $\pm$0.42\\
 NGC~6134 &	     6.94$\pm$0.11 &    8.63$\pm$0.07 &    0.069$\pm$0.008  &   0.109$\pm$0.004 &8.15 $\pm$0.38  &     7.91 $\pm$0.55\\
 NGC~6192 &	     7.07$\pm$0.08 &    9.00$\pm$0.05 &    0.128$\pm$0.003  &   0.120$\pm$0.003 &7.64 $\pm$0.00  &     8.89 $\pm$0.01\\
 NGC~6253 &	     5.37$\pm$0.19 &    7.95$\pm$0.09 &    0.210$\pm$0.007  &   0.194$\pm$0.012 &7.51 $\pm$0.23  &     6.76 $\pm$0.40\\
 NGC~6281 &	     8.03$\pm$0.03 &    8.31$\pm$0.08 &    0.055$\pm$0.001  &   0.017$\pm$0.003 &8.25 $\pm$0.08  &     8.04 $\pm$0.02\\
 NGC~6404 &	     6.76$\pm$0.11 &    7.05$\pm$0.12 &    0.078$\pm$0.007  &   0.021$\pm$0.001 &7.01 $\pm$0.11  &     6.88 $\pm$0.04\\
 NGC~6583 &	     5.91$\pm$0.15 &    6.83$\pm$0.12 &    0.131$\pm$0.004  &   0.072$\pm$0.004 &6.28 $\pm$0.21  &     6.38 $\pm$0.36\\
 NGC~6705 &	     4.91$\pm$0.18 &    8.87$\pm$0.11 &    0.364$\pm$0.018  &   0.288$\pm$0.011 &8.74 $\pm$0.03  &     6.63 $\pm$0.62\\
\hline
\hline
 & & & &  \\
 Cr~261   &	     4.86$\pm$0.23  &  10.61$\pm$0.05  &   0.261$\pm$0.011  &   0.372$\pm$0.022 &6.36 $\pm$0.07  &     8.69 $\pm$0.61\\
 IC~4651  &	     7.21$\pm$0.08  &   9.03$\pm$0.07  &   0.168$\pm$0.004  &   0.112$\pm$0.001 &8.49 $\pm$0.46  &     8.10 $\pm$0.58\\
 NGC~3960 &	     6.64$\pm$0.13  &  10.83$\pm$0.02  &   0.247$\pm$0.012  &   0.240$\pm$0.008 &9.25 $\pm$0.53  &    10.53 $\pm$0.26\\
 NGC~5822 &	     7.02$\pm$0.11  &   9.44$\pm$0.06  &   0.050$\pm$0.002  &   0.147$\pm$0.005 &8.66 $\pm$0.52  &     7.50 $\pm$0.46\\
 NGC~6134 &	     6.92$\pm$0.11  &   8.67$\pm$0.07  &   0.069$\pm$0.008  &   0.112$\pm$0.004 &8.22 $\pm$0.34  &     7.98 $\pm$0.53\\
 NGC~6192 &	     7.09$\pm$0.08  &   9.02$\pm$0.05  &   0.128$\pm$0.003  &   0.120$\pm$0.003 &7.62 $\pm$0.01  &     8.82 $\pm$0.01\\
 NGC~6253 &	     5.34$\pm$0.19  &   8.09$\pm$0.08  &   0.211$\pm$0.007  &   0.205$\pm$0.012 &5.43 $\pm$0.16  &     6.72 $\pm$0.89\\
 NGC~6281 &	     8.03$\pm$0.03  &   8.32$\pm$0.08  &   0.055$\pm$0.001  &   0.018$\pm$0.003 &8.26 $\pm$0.08  &     8.03 $\pm$0.03\\
 NGC~6404 &	     6.68$\pm$0.13  &   7.14$\pm$0.10  &   0.078$\pm$0.007  &   0.033$\pm$0.003 &7.09 $\pm$0.05  &     6.90 $\pm$0.02\\
 NGC~6583 &	     5.86$\pm$0.21  &   7.06$\pm$0.09  &   0.132$\pm$0.004  &   0.093$\pm$0.024 &6.21 $\pm$0.28  &     6.49 $\pm$0.19\\
 NGC~6705 &	     4.92$\pm$0.17  &   9.02$\pm$0.07  &   0.366$\pm$0.018  &   0.295$\pm$0.013 &8.75 $\pm$0.06  &     6.92 $\pm$0.48\\
\hline\hline
\end{tabular}
\caption{Output of the orbit calculation for the bar-less Allen \& Santillan model (upper section of the table)
and the n=2 Ferrers bar model (lower section of the table).}
\label{tab_orbit}
\end{table}
\end{landscape}

\begin{acknowledgements}
We thank an anonymous referee for her/his careful reading and useful comments. 
We thank Silvia Pedicelli for her help with the Cepheid data-set. 
LM is supported by the PRIN "Astroarcheologia galattica: la via locale per la cosmologia", P.I. Francesca Matteucci.
MZ is supported by the FONDAP Center for Astrophysics 15010003, the
BASAL Center for Astrophysics and Associated Technologies ATA PFB-06, Fondecyt Regular 1085278 and the MIDEPLAN Milky Way Millennium Nucleus P07-021-F.
\end{acknowledgements}


\begin{thebibliography}{}


\bibitem[1991]{allen91} Allen, C., \& Santillan, A.\ 1991, Revista Mexicana de Astronomia y Astrofisica, 22, 255 
\bibitem[2008]{allen08} Allen, C., Moreno, E., \& Pichardo, B.\ 2008, \apj, 674, 237 
\bibitem[1999]{alonso99} Alonso, A., Arribas, S., \& Mart{\'{\i}}nez-Roger, C.\ 1999, \aaps, 140, 261 
\bibitem[1989]{anders89} Anders, E., \& Grevesse, N.\ 1989, \gca, 53, 197 
\bibitem[2002]{an02} Andrievsky  S.  et al., 2002, \aap, 381, 32 
%\bibitem[2009]{ab09} Aumer, M., \& Binney, J.~J.\ 2009, \mnras, 397, 1286 
\bibitem[2004]{bn04} Balaguer-N{\'u}{\~n}ez, L., Jordi, C., Galad{\'{\i}}-Enr{\'{\i}}quez, D., \& Zhao, J.~L.\ 2004, \aap, 426, 819 
\bibitem[2000]{barklem00} Barklem, P.~S., Piskunov, N., \& O'Mara, B.~J.\ 2000, \aaps, 142, 467 
\bibitem[2010]{bellini10} Bellini, A., Bedin, L.~R., Pichardo, B., Moreno, E., Allen, C., Piotto, G., \& Anderson, J.\ 2010, \aap, 513, A51 
\bibitem[2010]{bensby10} Bensby T., Alves-Brito A., Oey, M.S., Yong D., Mel\'endez J. 2010, ApJL in press (astro-ph/1004.2833).
\bibitem[2008]{binney08} Binney, J., \& Tremaine, S.\ 2008, Galactic Dynamics: Second Edition, by James Binney and Scott Tremaine.~ISBN 978-0-691-13026-2 (HB).~Published by Princeton University Press, Princeton, NJ USA, 2008.,  
%\bibitem[1999]{boisser99} Boissier \& Prantzos 1999, \mnras, 307, 857
\bibitem[2006]{bra06} Bragaglia, A., \& Tosi, M.\ 2006, \aj, 131, 1544 
\bibitem[1994]{carraro94} Carraro, G., \& Chiosi, C.\ 1994, \aap, 288, 751 
\bibitem[1998]{carraro98} Carraro et al. 1998, \mnras, 296, 1045
\bibitem[2004]{carraro04} Carraro et al. 2004, AJ, 128, 1676
\bibitem[2005]{carraro05} Carraro, G., M{\'e}ndez, R.~A., \& Costa, E.\ 2005, \mnras, 356, 647 
\bibitem[2006]{carraro06} Carraro, G., Villanova, S., Demarque, P., McSwain, M.~V., Piotto, G., \& Bedin, L.~R.\ 2006, \apj, 643, 1151 
%\bibitem[2000]{carretta00} Carretta, E., Bragaglia, A., Tosi, M., \& Marconi, G.\ 2000, Stellar Clusters and Associations: Convection, Rotation, and Dynamos, 198, 273 
\bibitem[2004]{carretta04} Carretta et al. 2004, A\&A, 422, 951
\bibitem[2003]{chen03} Chen et al. 2003, AJ, 125, 1397
\bibitem[2001]{chiappini01} Chiappini et al. 2001, ApJ, 554, 1044
\bibitem[2006]{claria06} Clari{\'a}, J.~J., Mermilliod, J.-C., Piatti, A.~E., \& Parisi, M.~C.\ 2006, \aap, 453, 91 
\bibitem[2009]{colavitti09} Colavitti  E., Cescutti  G., Matteucci  F., Murante  G., 2009, \aap, 496, 429 
\bibitem[2004]{daflon04} Daflon, S., \& Cunha, K. 2004, ApJ, 617, 1115
\bibitem[2009]{davies09} Davies, B., Origlia, L., Kudritzki, R.-P., Figer, D.~F., Rich, R.~M., Najarro, F., Negueruela, I., \& Clark, J.~S.\ 2009, \apj, 696, 2014 
\bibitem[2000]{deha00} Deharveng, L., Pe{\~n}a, M., Caplan, J., \& Costero, R.\ 2000, \mnras, 311, 329 
\bibitem[1987]{fitz87} Fitzpatrick, M.~J., \& Sneden, C.\ 1987, \baas, 19, 1129 
\bibitem[1993]{friel93} Friel \& Janes 1993, A\&A, 267, 75
\bibitem[1995]{friel95} Friel 1995, ARA\&A, 33, 381
\bibitem[2002]{friel02} Friel et al. 2002, AJ, 124, 2693
\bibitem[2006]{friel06} Friel, E. D. 2006, in ``Chemical Abundances and Mixing in Stars in the Milky Way and its Satellites'', eds. S. Randich \& L. Pasquini, ESO Astrophysic Symposia, Vol.~24, Springer-Verlag, p.~3
\bibitem[2010]{friel10} Friel, E.~D., Jacobson, H.~R., \& Pilachowski, C.~A.\ 2010, \aj, 139, 1942
\bibitem[2009]{fu09} Fu, J., Hou, J.~L., Yin, J., \& Chang, R.~X.\ 2009, \apj, 696, 668 
\bibitem[2000]{girardi00} Girardi, L., Bressan, A., Bertelli, G., \& Chiosi, C.\ 2000, \aaps, 141, 371 
\bibitem[2000]{gonzales00} Gonzales \& Wallerstein \ 2000, PASP, 112, 1081
\bibitem[2003]{gratton03} Gratton, R.~G., Carretta, E., Claudi, R., Lucatello, S., \& Barbieri, M.\ 2003, \aap, 404, 187 
%\bibitem[2000]{hou00} Hou et al. 2000, A\&A, 362, 921
\bibitem[1983]{kilambi83} Kilambi, G.~C., \& Fitzgerald, M.~P.\ 1983, Bulletin of the Astronomical Society of India, 11, 226 
\bibitem[1987]{king87} King, D.~J.\ 1987, The Observatory, 107, 107 
\bibitem[1993]{kurucz93} Kurucz, R.~L.\ 1993, VizieR Online Data Catalog, 6039, 0 
\bibitem[1991]{kj91} Kjeldsen, H., \& Frandsen, S.\ 1991, \aaps, 87, 119 
\bibitem[2009]{ja09} Jacobson, H.~R., Friel, E.~D., \& Pilachowski, C.~A.\ 2009, \aj, 137, 4753 
\bibitem[1987]{js87} Johnson, D.~R.~H., \& Soderblom, D.~R.\ 1987, \aj, 93, 864 
\bibitem[1979]{janes79} Janes 1979, ApJS, 39, 135 
\bibitem[2003]{lepine01} L{\'e}pine, J.~R.~D., Acharova, I.~A., \& Mishurov, Y.~N.\ 2003, \apj, 589, 210 
\bibitem[2006]{levesque06} Levesque, E.~M., Massey, P., Olsen, K.~A.~G., Plez, B., Meynet, G., \& Maeder, A.\ 2006, \apj, 645, 1102 
\bibitem[2001]{loktin01} Loktin, A.~V., Gerasimenko, T.~P., \& Malysheva, L.~K.\ 2001, Astronomical and Astrophysical Transactions, 20, 607 
\bibitem[1989]{luck89} Luck, R.~E., \& Bond, H.~E.\ 1989, \apjs, 71, 559 
\bibitem[2006]{luck06} Luck  R., Kovtyukh  V., Andrievsky  S., 2006, AJ, 132, 902  
\bibitem[2007]{maciel07} Maciel, W.~J., Quireza, C., \& Costa, R.~D.~D.\ 2007, \aap, 463, L13 
\bibitem[2009]{m09} Magrini, L., Sestito, P., Randich, S., \& Galli, D.\ 2009, \aap, 494, 95 
\bibitem[2010]{pace10} Pace, G., Danziger, J., Carraro, G., Melendez, J., Francois, P., Matteucci, F., \& Santos, N.~C.\ 2010, arXiv:1002.2547 
\bibitem[2010]{pancino10} Pancino, E., Carrera, R., Rossetti, E., \& Gallart, C.\ 2010, \aap, 511, A56 
\bibitem[2003]{paulson03} Paulson, D.~B., Sneden, C., \& Cochran, W.~D.\ 2003, \aj, 125, 3185 
\bibitem[2003]{paunzen03} Paunzen, E., Maitzen, H.~M., Rakos, K.~D., \& Schombert, J.\ 2003, \aap, 403, 937 
\bibitem[2009]{pedicelli09} Pedicelli, S., et al.\ 2009, \aap, 504, 81 
\bibitem[1984]{pf84} Pfenniger, D.\ 1984, \aap, 134, 373 
\bibitem[1994]{phelps94} Phelps, R. L., Janes, K. A., Montgomery, K. A. 1994, \aj, 107, 1079
\bibitem[1998]{piatti98} Piatti, A.~E., Clari{\'a}, J.~J., Bica, E., Geisler, D., \& Minniti, D.\ 1998, \aj, 116, 801 
\bibitem[2004]{pic04} Pichardo, B., Martos, M., \& Moreno, E.\ 2004, \apj, 609, 144 
%\bibitem[2007]{py07} Pilyugin, L.~S., Thuan, T.~X., \& V{\'{\i}}lchez, J.~M.\ 2007, \mnras, 376, 353 
%\bibitem[1999]{portinari99} Portinari \& Chiosi 1999,A\&A, 350, 829
\bibitem[1992]{press92}Press, W.~H., Teukolsky, S.~A., Vetterling, W.~T., 
\& Flannery, B.~P.\ 1992, Cambridge: University Press, |c1992, 2nd ed.
\bibitem[2006]{randich06} Randich, S., Sestito, P., Primas, F., Pallavicini, R., \& Pasquini, L.\ 2006, \aap, 450, 557 
\bibitem[2007]{rood07} Rood, R.~T., Quireza, C., Bania, T.~M., Balser, D.~S., 
\& Maciel, W.~J.\ 2007, From Stars to Galaxies: Building the Pieces to Build Up the Universe, 374, 169 
\bibitem[2004]{salaris04} Salaris, M., Weiss, A., Percival, S. M. 2004, \aap, 414, 163
\bibitem[2010]{sanchez10} S{\'a}nchez, N., Vicente, B., \& Alfaro, E.~J.\ 2010, \aap, 510, A78 
\bibitem[1971]{sanders71} Sanders, W.~L.\ 1971, \aap, 14, 226 
%\bibitem[2009]{sc09} Sch{\"o}nrich, R., \& Binney, J.\ 2009, \mnras, 396, 203 
\bibitem[2010]{sc10} Sch{\"o}nrich, R., Binney, J., \& Dehnen, W.\ 2010, \mnras, 403, 1829 
\bibitem[2006]{sestito06} Sestito et al. 2006, A\&A, 458, 121
\bibitem[2007]{sestito07} Sestito, P., Randich, S., \& Bragaglia, A.\ 2007, \aap, 465, 185
\bibitem[2008]{sestito08} Sestito et al. 2008, A\&A, 488, 943
\bibitem[2006]{skrutskie06} Skrutskie, M.~F., et al.\ 2006, \aj, 131, 1163 
\bibitem[2009]{sm09} Smiljanic, R., Gauderon, R., North, P., Barbuy, B., Charbonnel, C., \& Mowlavi, N.\ 2009, \aap, 502, 267 
\bibitem[2010]{stanghellini10} Stanghellini, L., \& Haywood, M.\ 2010,  \apj, 714, 1096 
\bibitem[1997]{twarog97} Twarog et al. 1997, AJ, 114, 2556
\bibitem[2003]{twarog03} Twarog, B.~A., Anthony-Twarog, B.~J., \& De Lee, N.\ 2003, \aj, 125, 1383 
\bibitem[1955]{unsold55} Uns\"{o}ld, A.\ 1955, Berlin, Springer, 1955.~2.~Aufl.  
\bibitem[2010]{villanova10} Villanova, S., Randich, S., Geisler, D., Carraro, G., \& Costa, E.\ 2010, \aap, 509, A102 
\bibitem[2002]{wu02} Wu, Z.~Y., Tian, K.~P., Balaguer-N{\'u}{\~n}ez, L., Jordi, C., Zhao, L., \& Guibert, J.\ 2002, \aap, 381, 464 
\bibitem[2009]{wu09} Wu, Z.-Y., Zhou, X., Ma, J., \& Du, C.-H.\ 2009, \mnras, 399, 2146 
\bibitem[2005]{yong05} Yong et al. 2005, AJ, 530, 597
\bibitem[2010]{zac10} Zacharias, N., et al.\ 2010, \aj, 139, 2184 
\bibitem[1990]{zh90} Zhao, J.~L., \& He, Y.~P.\ 1990, \aap, 237, 54 




\end{thebibliography}
\end{document}